\makeatletter
\declare@file@substitution{revtex4-1.cls}{revtex4-2.cls}
\makeatother
\documentclass[manuscript]{aastex631}
\renewcommand{\sun}[0]{\odot}
\newcommand*\diff{\mathop{}\!\mathrm{d}}
\newcommand{\avg}[1]{\left< #1 \right>} 
\newcommand{\abs}[1]{\left| #1 \right|}
\newcommand{\rev}[1]{\textcolor{black}{#1}}
\begin{document}
	
\title{Velocity and Density Fluctuations in the Quiet Sun Corona}

\correspondingauthor{Michael Hahn}
\email{mhahn@astro.columbia.edu}

\author{Michael Hahn}
\affiliation{Columbia Astrophysics Laboratory, Columbia University, 550 West 120th Street, New York, NY 10027}

\author{Xiangrong Fu}
\affiliation{Los Alamos National Laboratory, P.O. Box 1663, Los Alamos, NM 87545}
\affiliation{New Mexico Consortium, 4200 W Jemez Rd \#200, Los Alamos, NM 87544}

\author{Stefan J. Hofmeister}
\affiliation{Columbia Astrophysics Laboratory, Columbia University, 550 West 120th Street, New York, NY 10027}

\author{Yifan Huang}
\affiliation{Los Alamos National Laboratory, P.O. Box 1663, Los Alamos, NM 87545}

\author{Alexandros Koukras}
\affiliation{Columbia Astrophysics Laboratory, Columbia University, 550 West 120th Street, New York, NY 10027}

\author{Daniel Wolf Savin}
\affiliation{Columbia Astrophysics Laboratory, Columbia University, 550 West 120th Street, New York, NY 10027}

\begin{abstract}

We investigate the properties and relationship between Doppler-velocity fluctuations and intensity fluctuations in the off-limb quiet Sun corona. These are expected to reflect the properties of Alfv\'enic and compressive waves, respectively. The data come from the Coronal Multichannel Polarimeter (COMP). These data were studied using spectral methods to estimate the power spectra, amplitudes, perpendicular correlation lengths, phases, trajectories, dispersion relations, and propagation speeds of both types of fluctuations. We find that most velocity fluctuations are due to Alfv\'enic waves, but that intensity fluctuations come from a variety of sources, likely including fast and slow mode waves, as well as aperiodic variations. The relation between the velocity and intensity fluctuations differs depending on the underlying coronal structure. On short closed loops, the velocity and intensity fluctuations have similar power spectra and speeds. In contrast, on longer nearly radial trajectories, the velocity and intensity fluctuations have different power spectra, with the velocity fluctuations propagating at much faster speeds than the intensity fluctuations. Considering the temperature sensitivity of COMP, these longer structures are more likely to be closed fields lines of the quiet Sun rather than cooler open field lines. That is, we find the character of the interactions of Alfv\'enic waves and density fluctuations depends on the length of the magnetic loop on which they are traveling. 
	
\end{abstract}

\section{Introduction} \label{sec:intro}

Wave-driven models of coronal heating consider that energy is transported to the corona by Alfv\'en waves, which damp in the corona and transfer that energy into particle heating. In most models, wave dissipation is mediated by a turbulent cascade of wave energy from the long length scales of Alfv\'en waves to smaller scales where particle interactions can heat the plasma. Turbulence arises through a nonlinear interaction between counter-propagating Alfv\'en waves \citep{Howes:PoP:2013}. In closed loops, counter-propagating waves arise naturally as waves excited at the two footpoints of the loop interact near the loop top. In open structures, such as coronal holes, the origin of sunward propagating waves is less clear. Large scale gradients in the magnetic field and density in the corona are thought to be too weak to efficiently reflect Alfv\'en waves \citep{Reville:ApJ:2018, Asgari:ApJ:2021}. 

The reflection of waves may be enhanced by interaction with fluctuations in the plasma density, such as an interaction with slow mode waves \citep{Asgari:ApJ:2021}. Alfv\'en waves are expected to be reflected from parallel gradients in the Alfv\'en speed and density fluctuations increase these gradients \citep{Heinemann1980, Musielak:PoFB:1992, Velli1993}. Theoretical models for open magnetic field regions show that adding density fluctuations to the plasma significantly enhances wave reflection and turbulent heating compared to a smooth radial variation of the background plasma parameters \citep{Asgari:ApJ:2021}. 

Another process that may produce the sunward propagating waves is the parametric decay instability (PDI). This is a nonlinear interaction in which a forward propagating ``pump'' Alfv\'en wave excites a forward acoustic wave and a backward secondary Alfv\'en wave. PDI is able to enhance turbulence and promote coronal heating in several ways. First, PDI directly generates a backward propagating Alfv\'en wave, which can interact with forward propagating Alfv\'en waves excited at the Sun to drive turbulence. Second, PDI produces large amplitude density fluctuations in the form of acoustic waves, which promotes Alfv\'en wave reflection. There has been much numerical work showing how PDI may lead to efficient coronal heating \citep{Shoda:ApJ:2016, Reville:ApJ:2018, Shoda:ApJ:2018, Shoda:ApJ:2019}. There is also evidence suggesting that PDI occurs in the transition region \citep{Hahn:ApJ:2022} and indirect evidence that it has influenced the solar wind \citep{Bowen:ApJ:2018, Kasper:PRL:2021}, although direct evidence for PDI in the solar wind is still lacking \citep{Zank:ApJ:2022, Zhao:ApJ:2022}. 

Besides these, there are many other potential interactions between Alfv\'en waves and density fluctuations. For example, low frequency large amplitude waves may drive density fluctuations due to the wave magnetic pressure \citep{Hollweg:PRL:1971, Hollweg:JGR:1971}. Additionally, Alfv\'en waves propagating along magnetic flux tubes with strong gradient in the density or magnetic field strength perpendicular to the mean field may undergo nonlinear phase mixing, generating fast mode waves \citep{Nakariakov:SolPhys:1997}. 

Our objective here is to identify observational relationships between velocity and density fluctuations in the quiet Sun corona. We take a global view, observing the entire off-limb low corona up to about 1.3~$R_{\sun}$ during a typical time that does not appear to be affected by any remarkable events. We study spectroscopic data observing an Fe~\textsc{xiii} line, which has a formation temperature of about 1.8~MK. Given the high temperature of the line, we observe material from the quiet Sun and active regions, but not the cooler plasma associated with coronal holes and open field lines. Velocity fluctuations are observed through the Doppler shift of the emission lines and the density fluctuations are related to the intensity of the line. Throughout, we refer to fluctuations rather than waves as there may be other sources of variations. We expect, however, that many velocity fluctuations are due to Alfv\'enic waves and the density fluctuations to compressive waves. We use the term Alfv\'enic phenomenologically to refer to magnetized plasma waves propagating at about the Alfv\'en speed, as a more precise description of their nature, such as being kink or torsional modes, is difficult to ascertain observationally. We test the validity of our interpretations of the fluctuations as part of the analysis. Additionally, although our broad view is not limited exclusively to a particular physical process, we do test in more detail for PDI using methods developed by \citet{Hahn:ApJ:2022}, such as by looking for a frequency scaling between the power spectra of intensity and velocity fluctuations. 

The rest of this paper is organized as follows. In Section~\ref{sec:obs} we give a brief overview of the instrument and describe the observations. The analysis is presented throughout Section~\ref{sec:anal}. There, we aim to describe a broad range of properties of the observed density and velocity fluctuations. We begin with characteristics that can be obtained from time variation within a single pixel, then describe how we can track the wave propagation through the plane of sky, and finally derive properties of the waves that depend on both the space and time evolution of the fluctuations. In the course of the analysis, we describe relevant comparisons to other works and possible conclusions based on those measurements. We defer some interpretations that rely on multiple aspects of the analysis to the discussion in Section~\ref{sec:dis}. Section~\ref{sec:sum} summarizes the main results. 

\section{Instrument and Observations}\label{sec:obs}

We use data from the Coronal Multichannel Polarimeter \citep[COMP;][]{Tomczyk:SolPhys:2008}. COMP uses a tunable filter and polarimeter to make spectropolarimetric observations at three  wavelengths that span the Fe~\textsc{xiii} 10747~\AA\ emission line. At each of the three wavelengths, COMP produces an image of the off-limb solar corona in the four Stokes parameters, however for the applications to be discussed here, we are concerned only with the spectroscopic aspects of the data. These images span a radial distance of about 1.05--1.3~$R_{\sun}$ with a spatial resolution of approximately 4.5$^{\prime\prime}$ per pixel. We have used the Level-2 COMP data product, available through the Mauna Loa Solar Observatory, which provides the intensity and Doppler velocity of the line, as interpreted by assuming that the Fe~\textsc{xiii} 10747~\AA\ line has a Gaussian shape \citep[see][]{Sharma:NA:2023}. 

\begin{figure}[t]
	\centering \includegraphics[width=1.0\textwidth, trim={2cm 0 2cm 0}]{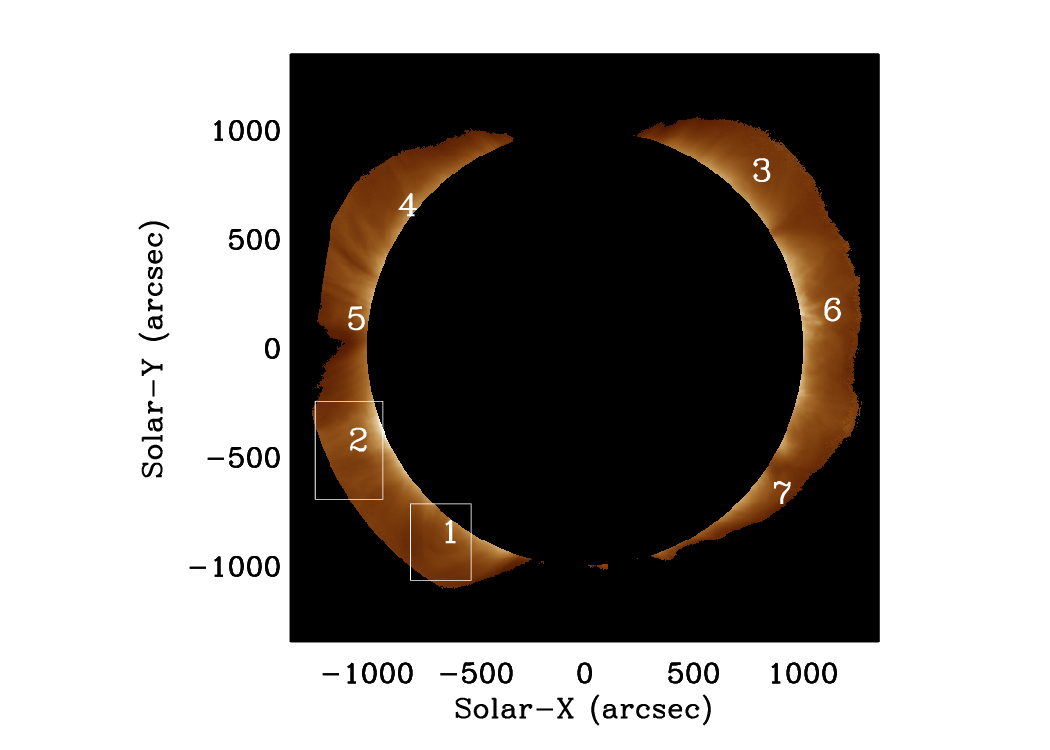}
	\caption{Context image showing the average Fe~\textsc{xiii} intensity observed by COMP during the observations. Labels highlight several apparent coronal structures: 1, 3, 4, and 7 label regions with relatively short closed loops; 2 and 5 indicate quiet Sun regions where the field lines are long and extend well beyond the field of view; in the vicinity of the label 6, there is an active region just behind the limb that leads to complex structures in the corona. The boxes surrounding 1 and 2 indicate the regions considered for the average power spectra discussed in Section~\ref{subsec:avgpow}.}
	\label{fig:context} 
\end{figure}

\begin{figure}[t]
	\centering \includegraphics[width=0.8\textwidth]{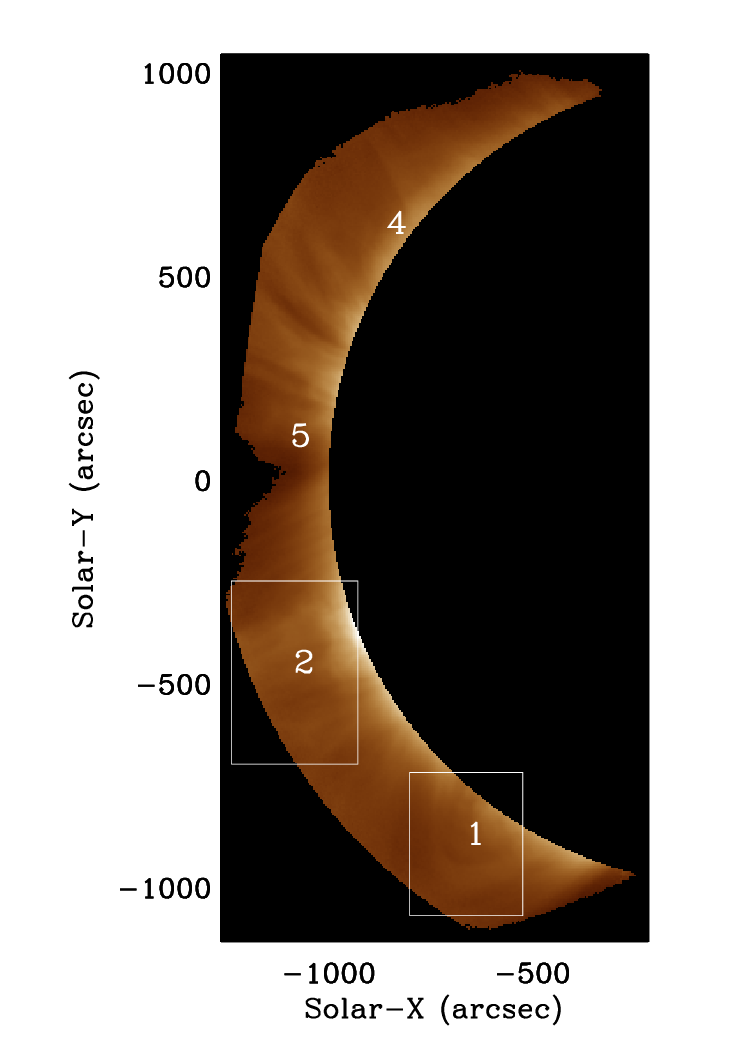}
	\caption{Same as Figure~\ref{fig:context}, but focusing on the East limb. }
	\label{fig:contexthalf} 
\end{figure}

Our data were obtained on 2016 July 25 from 20:25 to 21:53 UT. Figure~\ref{fig:context} presents a context image, showing the time-averaged Fe~\textsc{xiii} 10747~\AA\ intensity measured during the observation. The labels in the image are intended to highlight several apparent structures. Labels 1, 3, 4, and 7 are located near short closed loops, whose loop tops are clearly visible within the COMP field of view. Labels 2 and 5 are located near much longer structures, that appear nearly radial at the low heights within the field of view. These long structures are most likely the bases of long quiet Sun streamer loops. Although they appear ``open'' in this dataset, the Fe~\textsc{xiii} emission viewed by COMP is formed at high temperatures and so we do not expect to see the open magnetic fields from coronal holes. For future reference, we observationally define the short loops as those whose loop tops are below $1.3$~$R_{\sun}$ so that the entire loop is visible in the COMP data, and we define long loops as those that extend to much larger heights. 
Finally, on the west limb near the region labeled 6, there are several active regions (NOAA 12565, 12566, and 12567) behind the limb, resulting in complex magnetic structures in the corona. Because of the complex magnetic structure there and to improve the clarity of the presentation, we will often show figures focusing on the more quiescent East limb. A context image showing this more condensed field of view is shown in Figure~\ref{fig:contexthalf}. 

During the observation, COMP obtained intensity and velocity data with a regular cadence of $\Delta t = 30$~s. Our aim is to determine the typical relationships between velocity and density fluctuations in the corona. There was no particular scientific significance to the selected date, which was chosen at random. By coincidence, it turns out that the same dataset was one of those studied by \citet{Sharma:NA:2023} in their recent study of the perpendicular correlation length for Alfv\'enic waves. 

Our data set contains a time series of two-dimensional spatial maps of the intensity and Doppler shift in the corona. The COMP Level-2 data calibration pipeline sets to zero any pixels where the intensity was below a threshold of two millionths of the sky brightness. However, these pixels are not necessarily the same in each image. We therefore masked the dataset by omitting any pixel in all the maps where any of the 176 data points in the time series had an intensity that fell below that threshold. There are similar artifacts close to the occulted region, which we remove by masking out pixels below 1.08~$R_{\sun}$. These filters remove about 15\% of the initial data, or about 9000 out of 66000 pixels.  

For the analysis of fluctuations it is necessary to have data whose mean is zero and remove any long term trends. This was done for the velocity fluctuations, $\delta v$, by performing a linear fit to the time series at each pixel and then subtracting that linear trend. The intensity fluctuations were treated in the same way, except that we also normalized the data by the intensity trend so that we study $\delta I/I$, which we will denote as $\delta i$ throughout. 

For collisionally excited lines, the intensity is proportional to the square of the electron density $n_{\mathrm{e}}^2$ \citep[][]{Ultraviolet}. Consequently, density fluctuations and intensity fluctuations are related by $\delta n_{\mathrm{e}}/n_{\mathrm{e}} \sim \delta i/2$. We can apply this analysis to our study here because, although the level population of the Fe~\textsc{xiii} 10747~\AA\ line in the corona is primarily set by radiative decays from higher levels, those levels are collisionally excited. Hence the Fe~\textsc{xiii} line intensity fluctuations will exhibit the above relationship to density fluctuations. 

\section{Analysis and Results}\label{sec:anal}

\subsection{Amplitudes}\label{subsec:amp}

The power spectra and amplitudes of $\delta v$ and $\delta i$ can be inferred from the time series of data at each good pixel in the dataset. In principle, the root-mean-square (RMS) amplitude of the fluctuations could be determined by taking the square root of the variance of the time series data. However, this overestimates the amplitude due to the noise contribution from nonphysical sources of variation, such as detector noise, photon counting statistics, and variations due to stray light from other locations within the image. We applied a correction to mitigate some of these factors. The same analysis is used for both $\delta v$ and $\delta i$, but for brevity we discuss the analysis using $\delta v$ as an example. 

First, we computed the discrete Fourier power spectrum (periodogram) at each pixel for $\delta v(t)$. The resulting power spectrum $P_{\delta v}(f_{j})$ describes the variance in $v(t)$ due to oscillations at frequency $f_j$. The total $\sum_j P_{\delta v}(f_j)$ is equal to the variance of $\delta v(t)$, so the RMS amplitude of the fluctuations is the square root of the summed power spectrum. In order to reduce the statistical uncertainty in the estimated $P_{\delta v}(f_j)$, the power spectrum from a given pixel at position $(x,y)$ is averaged with that of its neighbors. This averaging produces a more precise estimate of the power spectrum, but does not remove the contributions of noise. 

In order to estimate the noise level, we assume that the physical sources of variation in $\delta v(t)$ are the same for the reference pixel and its neighbors and that variations from noise are random and uncorrelated so that we can decompose the measurements into a physical signal, $s(t)$, plus random noise, $e(t)$, giving $\delta v(t) = s(t)+e(t)$. Then, we can average the time series to obtain $\overline{\delta v(t)}$ whose variance from physical sources remains the same but for which the contributions to the variance from random noise are reduced by a factor of $1/K$, where $K \leq 9$ represents the number of pixels including the reference pixel and up to eight neighboring good pixels. This relation also holds at each frequency in the power spectrum, and we can estimate the noise power spectrum as $P_{e}(f_j) = (P_{\delta v}(f_j)-P_{\overline{\delta v}}(f_j))/(1-1/K)$. The inferred noise power spectrum, $P_{e}(f_j)$ did not exhibit a clear frequency-dependence, so to reduce statistical uncertainties, we further assume the noise level to be constant (i.e., white noise) and take the average $P_e$ over all the frequencies. Finally, we subtract this $P_e$ from $P_{\delta v}(f_j)$, sum the power over all the frequencies, and take the square root to obtain the RMS amplitude of the $\delta v$ fluctuations. The noise estimate reduces the average inferred amplitude of $\delta v$ fluctuations by about $0.1$~$\mathrm{km\,s^{-1}}$ ($\approx 15\%$) and reduces the inferred amplitudes of $\delta i$ fluctuations by about $0.004$ ($\approx 20\%$), compared in each case to when no correction is made. 

There are clearly some limitations to this method. First, we can only remove uncorrelated random sources of noise. Any noise sources that affects neighboring pixels in a similar way cannot be removed. For example, stray light fluctuations due to variations on the solar disk are likely similar in these off-limb pixels and their effects are not removed. Additionally, averaging the time series reduces the amplitude of any fluctuations that are out of phase with one another. Most relevant to this work, any traveling waves will have slightly different phases in neighboring pixels along the wave path and so the inferred amplitude will be slightly reduced by the averaging. 

\citet{Morton:ApJ:2016} used a different method to estimate the noise level from COMP data. They smoothed the data using a 3-point boxcar average over time and then subtracted the smoothed fluctuations from the original time series. The standard deviation of the residuals then gives an estimate of the noise level. We found that using their method compared to our approach produces $\delta v$ amplitudes that are smaller by on average $0.07$~$\mathrm{km\,s^{-1}}$ and $\delta i$ amplitudes that are smaller by about 0.001. That the amplitudes are systematically smaller may be due to the boxcar-average method implicitly assuming that all high frequency signal is due to noise. Nevertheless, these discrepancies are relatively small, $\approx 10\%$ of the amplitudes, and so the following results are not very sensitive to the way the noise level is estimated. 

\begin{figure}[h]
	\centering \includegraphics[width=0.8\textwidth]{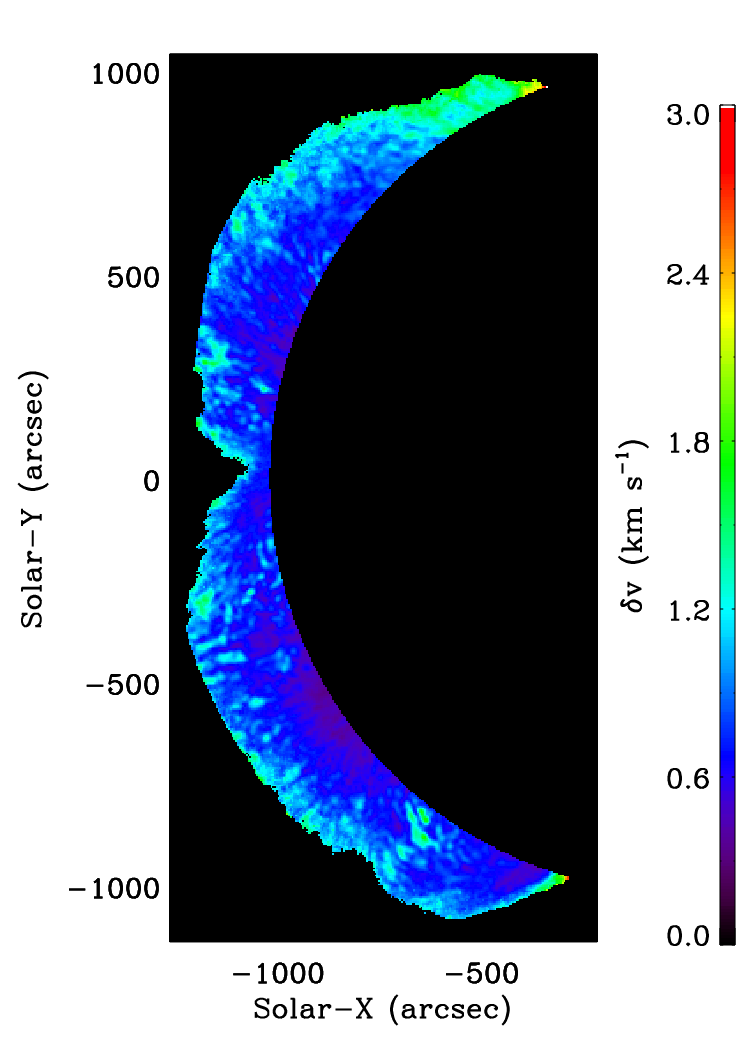}
	\caption{Amplitudes of $\delta v$ fluctuations.}
	\label{fig:ampv} 
\end{figure}

\begin{figure}[h]
	\centering \includegraphics[width=0.8\textwidth]{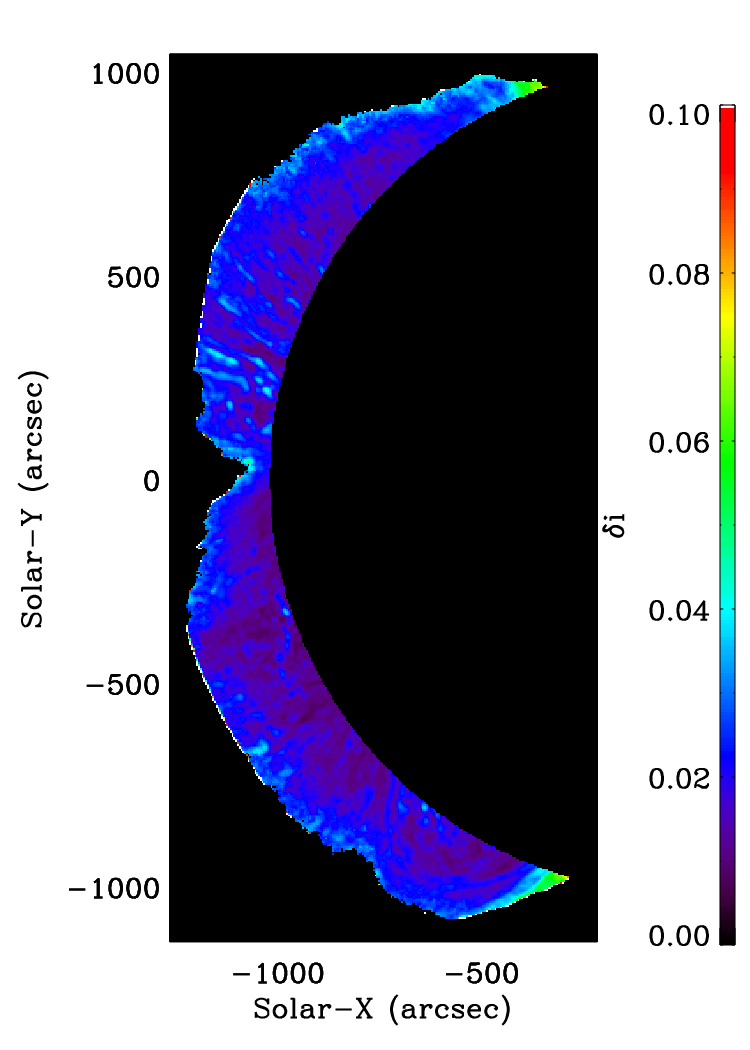}
	\caption{Same as Figure~\ref{fig:ampv}, but for $\delta i$ fluctuations.}
	\label{fig:ampi} 
\end{figure}

Figures~\ref{fig:ampv} and \ref{fig:ampi} show maps of the RMS amplitudes for the $\delta v$ and $\delta i$ fluctuations on the east limb, respectively. The $\delta v$ amplitudes are found to be in the range of about $0.5$--$1$~$\mathrm{km\,s^{-1}}$ throughout most of the field of view, which agrees with previous measurements based on the COMP data \citep[e.g.,][]{Tomczyk:Sci:2007, Tomczyk:ApJ:2009}. Many of the $\delta v$ fluctuations are due to Alfv\'enic waves, but these measurements likely underestimate the true wave amplitudes due to the implicit line-of-sight averaging \citep{McIntosh:ApJ:2012}. 

The $\delta v$ amplitudes appear to be larger near the poles. One possibility is that this represents a difference in the characteristics of the waves on the long field lines bordering the coronal holes near the poles as compared to the shorter loops found at lower latitudes. The high formation temperature for Fe~\textsc{xiii} makes it unlikely that any of the emission observed by COMP comes from open field regions. As such, it seems more likely that the increased amplitudes at the poles is due to line of sight effects, such as a change in the average viewing angle of the line of sight with respect to the magnetic field. We may be observing at the poles a larger contribution from $\delta v$ fluctuations in the parallel flows along the field rather than the transverse waves that likely predominate in the lower latitude parts of the observations. Alternatively, there may be less line-of-sight averaging near the poles because the depth of Fe~$\textsc{xiii}$ emitting material along the line of sight is smaller due to the presence of the cooler coronal hole material. As line-of-sight averaging is known to reduce the inferred amplitudes, having systematically less averaging at the poles would result in a systematically larger inferred amplitude. 

The $\delta i$ amplitudes in the low corona appear to be typically on the order of a few percent and increase systematically with height. However, these amplitudes are likely to be systematically underestimated at the largest heights, because we have not accounted for scattered light. The relative intensity fluctuation $\delta i$ is the intensity fluctuation $\delta I$ normalized by the time-averaged intensity at that position; but this average intensity has not been corrected for scattered light. Thus, the normalization factor is overestimated and the $\delta i$ amplitudes are likely underestimated. The contribution of scattered light to the total intensity is expected to grow with height, so that the underestimate in the $\delta i$ amplitudes also grows with height. 

The largest $\delta i$ amplitudes also appear to be associated with the tops of coronal loops. This is partly due to the general trend that $\delta i$ grows with height, but may also be due to turbulence. Previous observations by \citet{DeMoortel:ApJ:2014} showed that the power spectrum of $\delta v$ fluctuations was broader at loop tops and suggested that this was due to turbulence. Turbulence might also lead to an increase in density fluctuations. Density fluctuation amplitudes in open field regions have been observed to increase with height and that increase was associated with apparent damping of the Alfv\'enic waves \citep{Hahn:ApJ:2018}. 

\clearpage
\subsection{Power Spectra}\label{subsec:avgpow}

In order to compare the typical power spectra for $\delta v$ and $\delta i$ fluctuations, we have averaged the power spectra over the boxed regions in Figure~\ref{fig:context}: region 1 with short loops and region 2 with long radial structures. To perform this averaging, we computed the power spectra at each pixel within each region. Histograms of the statistical distribution of the measured power within a given frequency bin showed that the results appeared to follow an exponential distribution. This is the expected statistical distribution for the power at a given frequency \citep{Vaughan:AA:2005}. The reason is that the real and imaginary parts of the discrete Fourier transform are expected to follow a normal distribution and so their squared sum, the power, follows a chi-squared distribution with two degrees of freedom, which is an exponential distribution. We then take the average power derived from all of those pixels within each frequency bin. Finally, we subtracted the noise power level that was inferred using the boxcar smoothing method of \citet{Morton:ApJ:2016} discussed above in Section~\ref{subsec:amp}. This noise level was about $2\times10^{-3}$~$\mathrm{km^2\,s^{-2}}$ for the $\delta v$ fluctuations and $2 \times 10^{-6}$ for the $\delta i$ fluctuations. The exponential distribution has the property that the standard deviation is equal to the mean, so the uncertainty on each power spectrum is essentially 100\% \citep[see e.g.,][]{Vaughan:AA:2005}. However, the statistical uncertainty of the average power at a given frequency is the power at that frequency divided by $\sqrt{N}$, where $N$ is the number of samples. Here, $N=3778$ samples for region 1 and $N=5571$ samples for region 2, so the statistical uncertainties are very small. However, this does not account for pixel-to-pixel variations in the power spectrum, which is likely the dominant source of uncertainty in the final results. 

\begin{figure}[h]
	\centering \includegraphics[width=0.8\textwidth]{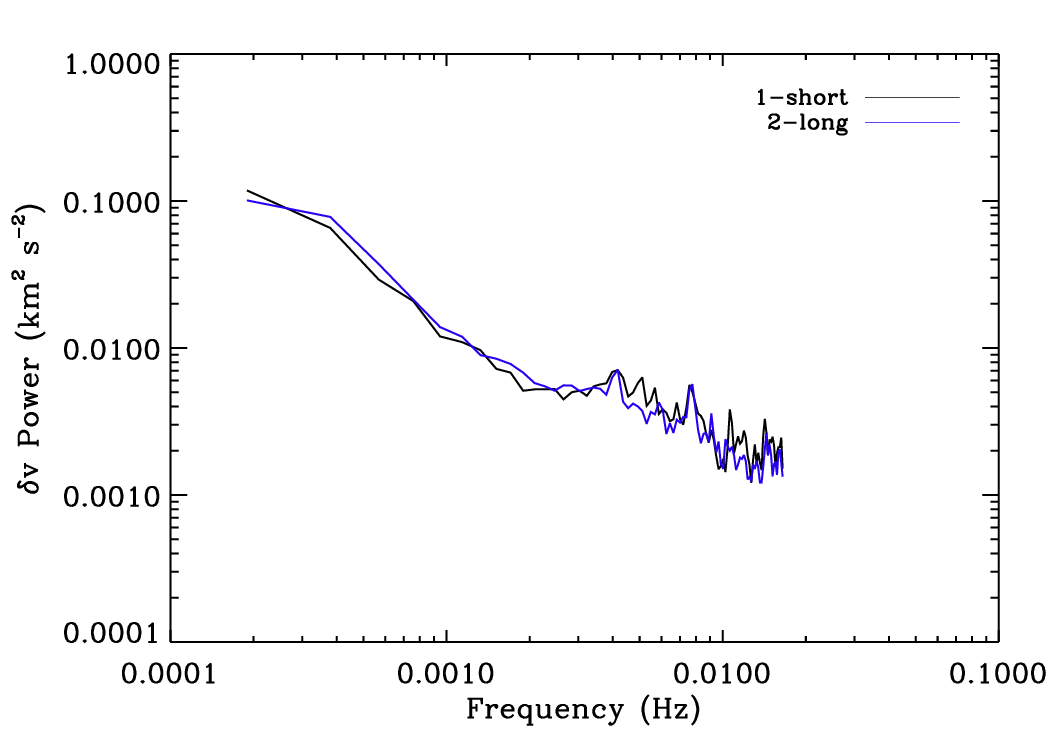}
	\caption{Average power spectra for $\delta v$ fluctuations in the regions labeled 1 and 2 in Figures~\ref{fig:context} and \ref{fig:contexthalf}, which correspond to short closed loops and long radial structures, respectively. }
	\label{fig:psdv} 
\end{figure}

Figure~\ref{fig:psdv} shows the average power spectra for the $\delta v$ fluctuations. The power spectrum for $\delta v$ consists of a power law continuum with a hump near $f \approx 3.5$~mHz. The power spectra from the short loops and the long radial structures are nearly identical, both in shape and in magnitude. These results are also very similar to previous COMP measurements \citep[e.g.,][]{Tomczyk:ApJ:2009, Morton:ApJ:2016}. 

\begin{figure}[h]
	\centering \includegraphics[width=0.8\textwidth]{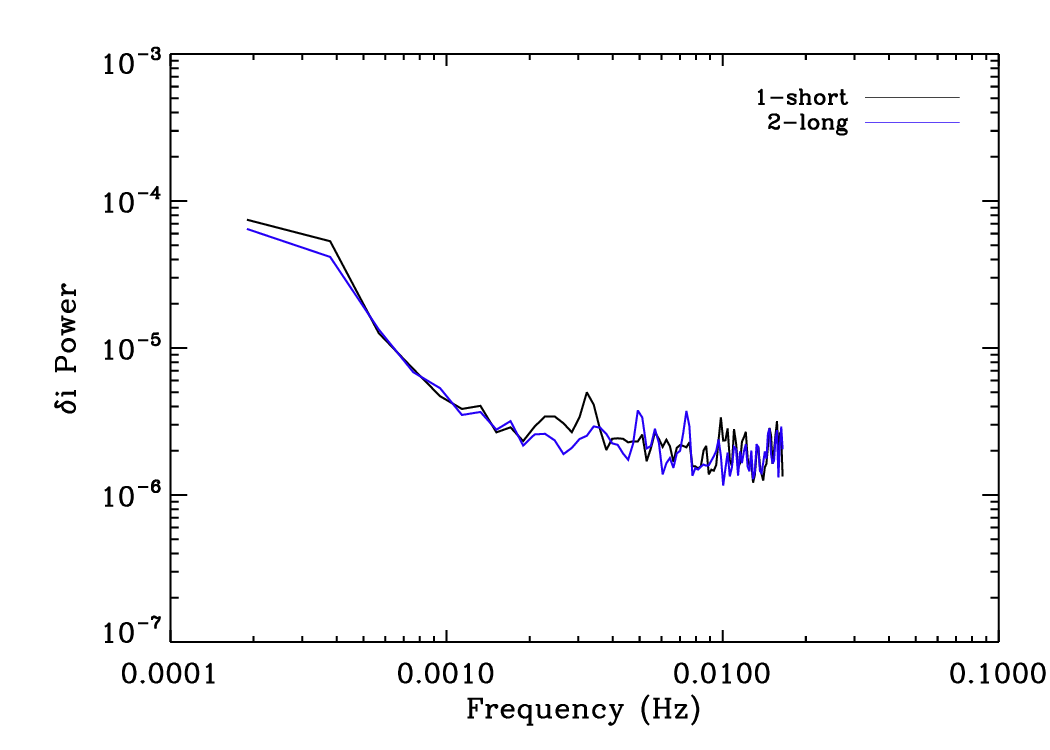}
	\caption{Same as Figure~\ref{fig:psdv}, but for $\delta i$ fluctuations. }
	\label{fig:psdi} 
\end{figure}

The power spectra for the $\delta i$ fluctuations are shown in Figure~\ref{fig:psdi}. Here, we find a rapid decay for low frequencies, that levels off to a nearly flat white noise type spectrum above about 2~mHz. One possible reason for this difference in the overall shape of the power spectrum between $\delta v$ versus $\delta i$ fluctuations is that the $\delta v$ fluctuations may be due largely to wave-like oscillations dominated by a few frequencies, whereas the $\delta i$ fluctuations are caused by many underlying physical processes, such as waves, heating, flows, etc., occurring on a variety of timescales. The overall shape of the $\delta i$ power spectrum is very similar between the short and long loops, except near 3~mHz. At that frequency, the short closed loops show a clear peak, whereas this peak is much less prominent in the region with the long radial structures. To verify this, we performed the same analysis using data in the vicinity of regions 4 and 5 in Figure~\ref{fig:context}, containing short versus long loops, respectively. Those regions also exhibited the relationship that the 3~mHz peak in the $\delta i$ fluctuations is broader and shallower on the long streamers. 

One possible interpretation of this result is that the $3$~mHz power represents the perturbations at the base of the corona that excite both Alfv\'enic waves and density fluctuations. On the short loops, power at this frequency is injected at both ends of the loop and the influence of the boundary remains evident throughout the loop. On the longer structures, the excited Alfv\'enic wave power evidently does not undergo a significant change, but the density fluctuation power spectrum evolves. This may occur through damping of the initially excited acoustic modes and the excitation of new density fluctuations with different characteristics at larger heights, such as through parametric decay of Alfv\'en waves or other processes. 

\subsection{Frequency scaling}\label{subsec:fac}

We next examine whether there is a frequency-scaling relationship between the $\delta v$ and $\delta i$ fluctuations. \citet{Hahn:ApJ:2022} found that the power spectra of $\delta v$ and $\delta i$ fluctuations in the transition region were related to one another by a scaling of the frequency axis. They considered this a signature of PDI, interpreting the frequency scaling factor to represent the ratio of the pump Alfv\'en wave frequency to the frequency of the PDI resonantly excited slow mode wave. Here, we used the same method to determine the scaling factor, $\alpha$, by which the frequency axis of the $\delta i$ power spectrum $P_{\delta i}(\alpha f)$ should be scaled to maximize the cross correlation with the $\delta v$ power spectrum $P_{\delta v}(f)$. For each pixel in the dataset, we have averaged the power spectra over all the neighboring good pixels, which is typically an average over $K=9$ pixels.

\begin{figure}[h]
	\centering \includegraphics[width=0.8\textwidth]{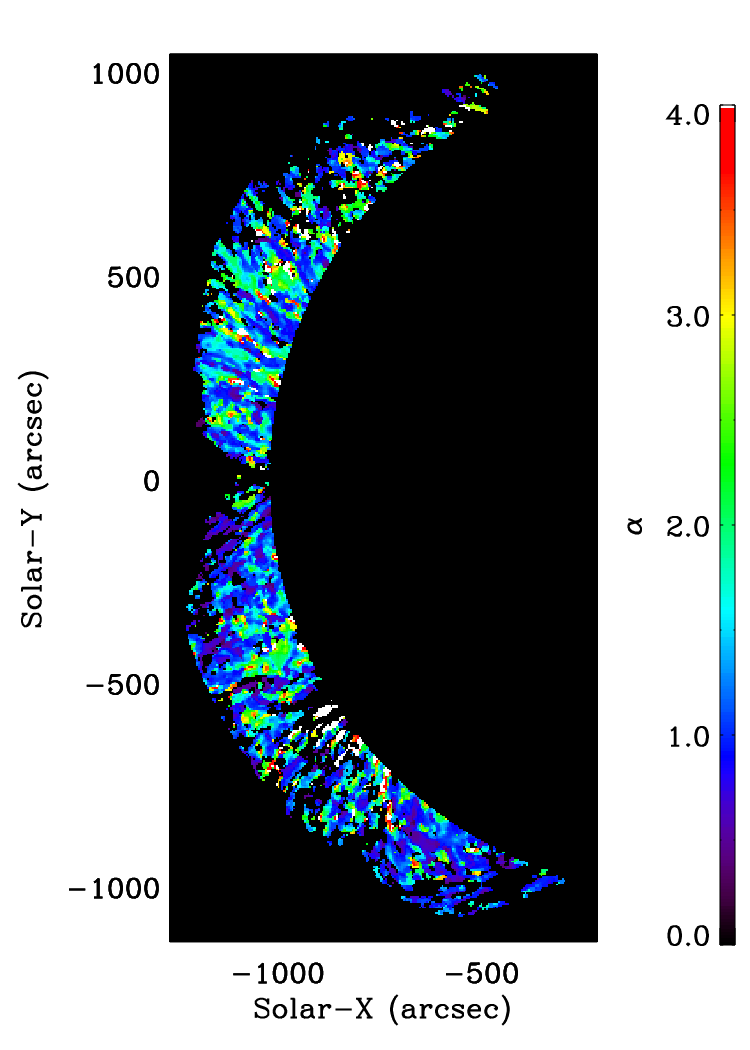}
	\caption{Frequency scaling factor between $\delta i$ and $\delta v$ fluctuations.}
	\label{fig:facmap} 
\end{figure}

Figure~\ref{fig:facmap} shows the scaling factors between the $\delta i$ and $\delta v$ fluctuations. Here we have suppressed any pixels where the maximum correlation coefficient was below $0.7$, which is about 40\% of the unmasked pixels. The cutoff of 0.7 was chosen because it corresponds to a $p$-value of about 0.05, i.e., the probability that the correlation arose by chance is estimated to be less than 5\%. However, that estimate relies on the idealized condition that the underlying correlated variables followed a normal distribution. Here, we consider it only as a way for removing distracting points from the plot of the scaling factor that most likely do not correspond to actual correlations. 

We do not find a clear relationship between intensity and velocity fluctuations. We commonly find $\alpha$ values close to unity, especially in the coherent closed loop structure seen in the lower left quadrant. One possibility is that these oscillations represent the velocity and compressive fluctuations associated with the same wave mode. 
Somewhat larger values of $\alpha \approx 1.5-2$ are seen on the long field lines at lower latitudes. This relationship might be consistent with PDI, discussion of which we defer to Section~\ref{sec:dis}. 

\clearpage 
\subsection{Phase Relationships}\label{subsec:phase}

The phase relationship between the $\delta v$ and $\delta i$ fluctuations also provides clues as to the nature of these fluctuations. If the fluctuations arise from the same underlying wave mode, we expect $\delta v$ and $\delta i$ to have a definite phase difference. For example, a fast sausage mode observed perpendicular to the magnetic field should exhibit a phase difference of $\pi/2$ between the velocity and density oscillations \citep{Gruszecki:AA:2012}, or a slow mode wave observed parallel to the field should exhibit a phase difference of zero \citep[e.g.,][]{Ofman:ApJ:2000}. 

We have examined the phase differences between the fluctuations by computing the Fourier cross-power spectrum between $\delta v$ and $\delta i$ in each pixel \citep{White:ApJ:1979}. The phase differences $\phi(x,y,f)$ are derived from the real and imaginary parts of the cross-power spectrum at each frequency and at each pixel. As discussed above, the Fourier coefficients have large uncertainties so an averaging is needed in order to assess whether the phase difference is significant. For this, we assume that the relation between the $\delta v$ and $\delta i$ fluctuations varies slowly in space so that the phase differences should be similar between a given pixel and its eight neighboring pixels. Then, we compute the average phase $\phi$ over the central pixel and up to eight good neighboring pixels and a quantity that measures the dispersion of these $\phi$ values. 

The analysis of the phase angle requires special treatment because $\phi$ is cyclical, so one cannot simply take the standard arithmetic mean. Instead, we find the mean by averaging the unit vectors corresponding to each value of $\phi$ \citep{Mardia:book}. We compute the mean value of $\bar{C} = \avg{\cos(\phi)}$ and $\bar{S} = \avg{\sin(\phi)}$, where the angle brackets denote the arithmetic mean. Then the average value of the phase difference is $\bar{\phi} = \arctan{(\bar{S}/\bar{C})}$, producing angle in the range of $-\pi$ to $\pi$. The concentration of the angles is described by the mean resultant length $\bar{R} = (\bar{C}^2+\bar{S}^2)^{1/2}$, which is a number between 0 and 1. That is, if all of the angles in the set over which the average is taken were in the same direction, then $\bar{R}=1$ and if they are all randomly distributed $\bar{R}=0$. It is useful to define instead the dispersion $\bar{V}=1-\bar{R}$, also called the circular variance, which corresponds to the more familiar interpretation that this number represents a spread of values so that a smaller number suggests a stronger relationship. 

\begin{figure}[h]
	\centering \includegraphics[width=0.8\textwidth]{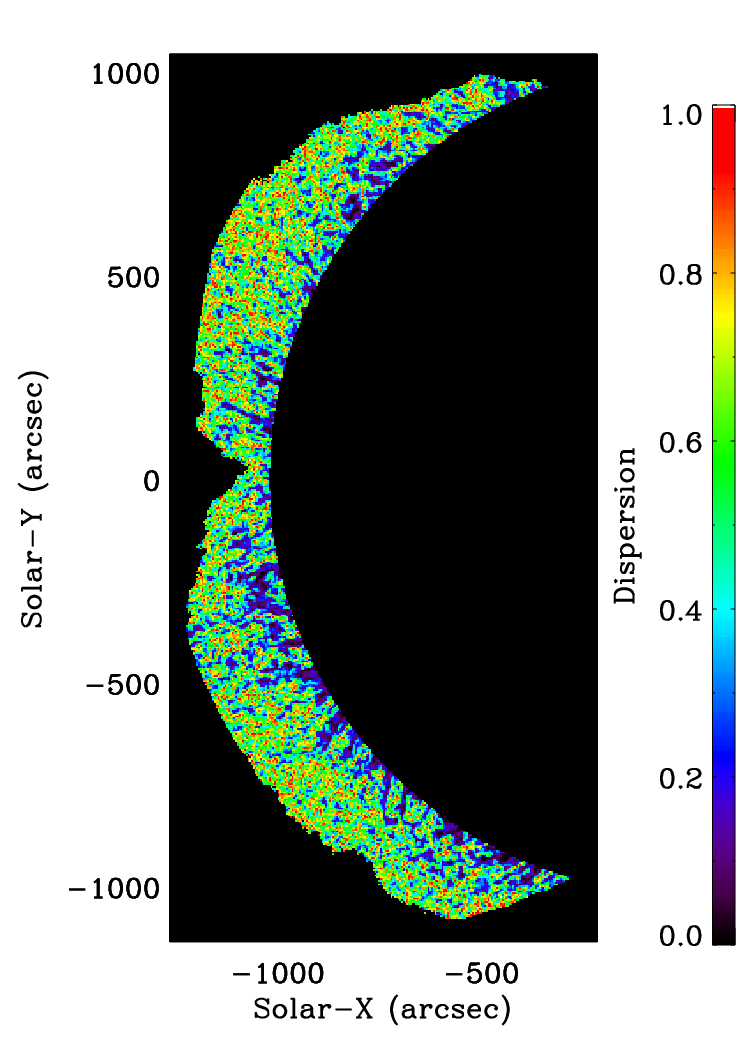}
	\caption{Dispersion, $\bar{V}$, of the phase differences between the $\delta v$ and $\delta i$ fluctuations at 3.5~mHz. A dispersion of $0$ indicates that the phases of fluctuations at neighboring pixels were the same, while a dispersion of $1$ indicates that the phases of neighboring pixels were all different.}
	\label{fig:vbar35} 
\end{figure}

We have applied the above steps to compute $\bar{\phi}(x,y,f)$ and $\bar{V}(x,y,f)$, where the average is taken over each pixel and the up to eight neighboring good pixels. Figure~\ref{fig:vbar35} presents a map of the dispersion, $\bar{V}$ at 3.5~mHz, corresponding to the peak in the $\delta v$ and $\delta i$ fluctuation power spectra. We find that there are low-dispersion structures extending out from the base of the corona that appear to largely follow the magnetic field lines. At larger heights the dispersion is large, corresponding to essentially random phases. This may be due to increasing scatter in the data from noise or to physical processes that cause the break up of coherent wave modes. Even at low heights, though, there are many locations with a high dispersion. The clear spatial structuring of the $\bar{V}$ that aligns with observed coronal structures supports that there are strong phase relationships between $\delta v$ and $\delta i$ fluctuations. 

\begin{figure}[h]
	\centering \includegraphics[width=0.8\textwidth]{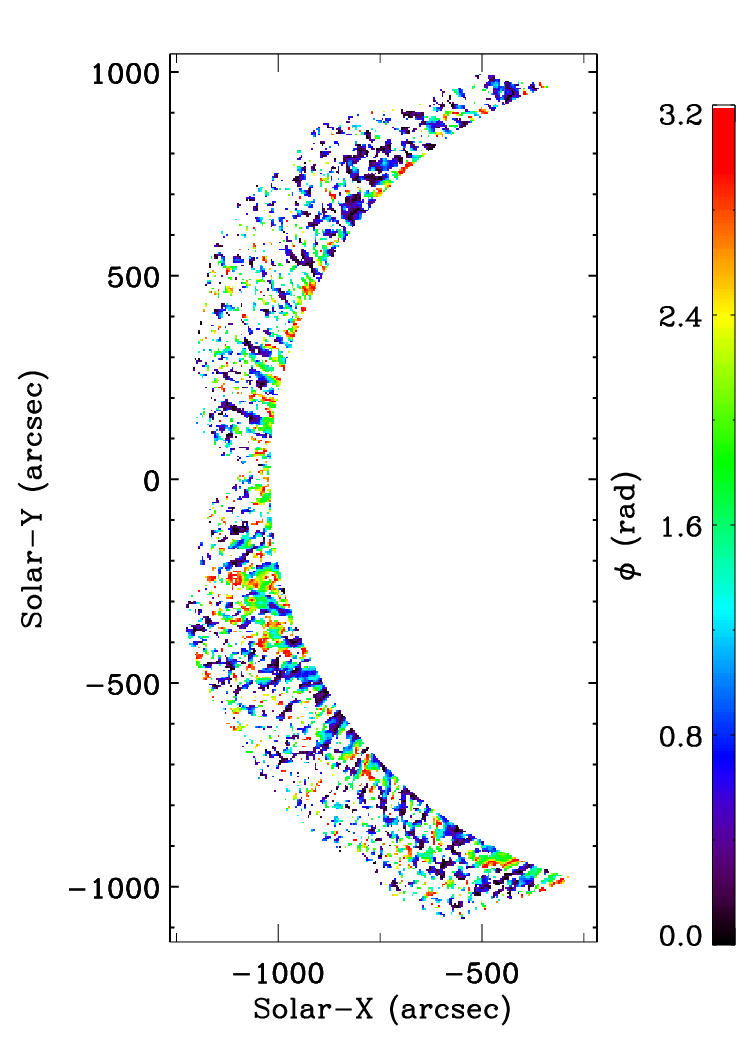}
	\caption{Nearest-neighbor-averaged phase differences $\bar{\phi}$ between $\delta v$ and $\delta i$ fluctuations at 3.5~mHz for locations with a statistically significant low dispersion.}
	\label{fig:phi35} 
\end{figure}

Figure~\ref{fig:phi35} shows the absolute value of the measured phase angles $\abs{\bar{\phi}(x,y)}$ at the same frequency of 3.5~mHz. The reason for plotting the absolute value is that $\phi=-\pi$ and $\phi = \pi$ correspond to the same phase difference. In this plot we have suppressed locations with a high dispersion by plotting only those pixels where $\bar{V} \leq 0.4$. This cutoff was chosen as it corresponds, approximately, to a $p$-value of 0.05 \citep{Mardia:book}. Although there is a great deal of scatter in the data, there appears to be a tendency for the phase values near the equator to have have $\phi \sim \pi/2$, whereas at middle to high latitudes $\phi \sim 0$ is more common. In terms of coronal structures the equatorial region contains more long field lines, whereas the middle to high latitudes have the shorter closed loops. 

\clearpage
\subsection{Wave tracking}\label{subsec:tracks}

For many properties of the waves, it is necessary to measure the fluctuations in both time and space along the path of the waves. So, a first step in such an analysis is to determine the trajectories of the waves. We used a method based on measuring the cross correlation of the fluctuations to determine the wave travel direction at each pixel. Once the directions are known, then the wave paths in the plane of the sky can be found by stepping along the trajectories using Euler's method. In the course of this analysis, we also determine the parallel and perpendicular correlation lengths. 

Our method for determining the wave direction is based on the correlation method presented by \citet{Tomczyk:Sci:2007}. Our measured power spectra (Figure~\ref{fig:psdv}), and previous work by others, has shown $\delta v$ fluctuations in the corona tend to have a peak power around 3.5~mHz and we found $\delta i$ fluctuations to have a peak at a similar low frequency. So, we first apply a low pass Butterworth filter to the data with a cutoff above 7~mHz to reduce the impact of high frequency noise. (Note that this filter was only applied for this wave tracking step and not in any of the following analyses of the power spectra along the wave path.) Then, for each pixel we compute the cross correlation $c_{ij}$ between that pixel $i$ and all of its neighbors $j$ within $\pm 12$ pixels ($\pm 54^{\prime\prime}$). This reveals an oval island of high cross correlation with a long axis oriented along the direction of wave travel. The $\pm 12$ pixel limit was selected because it was found to be large enough to contain these islands without being so large as to be influenced by random cross correlations with distant structures. To quantify the travel direction, we can find the slope of the line that will minimize the square of the perpendicular distance to each point in the island above a threshold high correlation. Here, we defined a high correlation to be a correlation coefficient $> 0.6$. This value was chosen by inspection as it results in islands with a large enough number of pixels to estimate the orientation consistently, while also providing a reasonable noise cutoff. The computation of the best fit line was performed using a principal component analysis. That is, we computed the covariance matrix for the coordinates, then found the eigenvalues and eigenvectors of the covariance matrix. The direction parallel to the wave path is given by the eigenvector corresponding to the largest eigenvalue and the other eigenvector gives the perpendicular direction. 

\begin{figure}[h]
	\centering \includegraphics[width=1.0\textwidth, trim={2cm 0 2cm 0}]{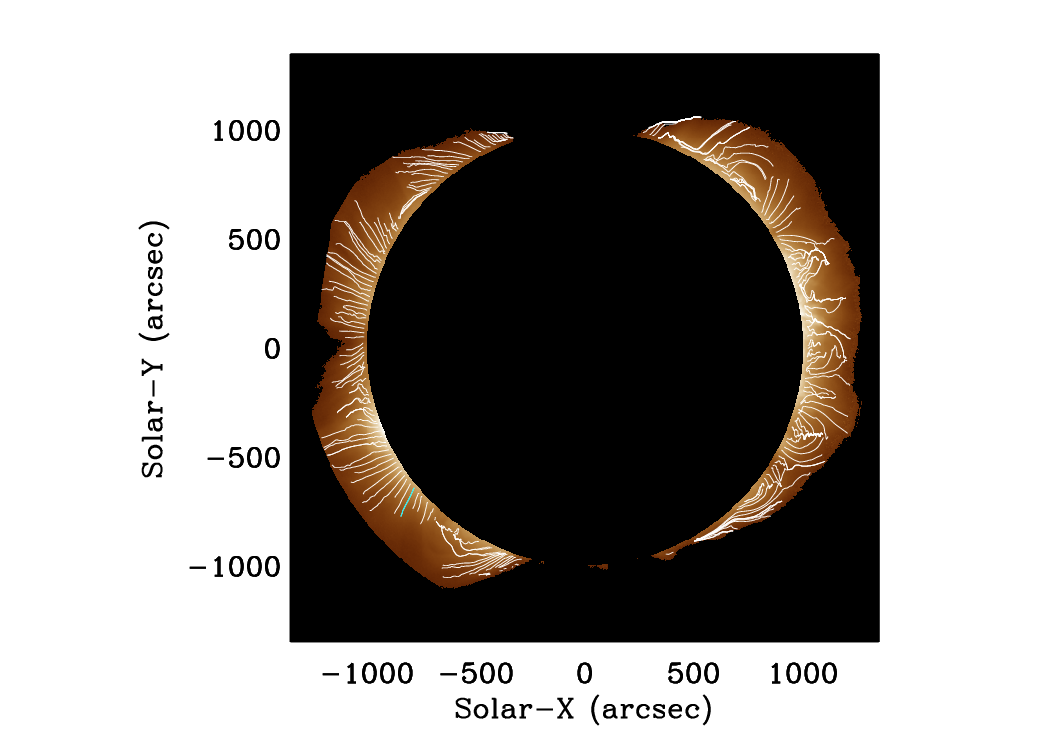}
	\caption{Wavetracks derived from the $\delta v$ fluctuations. Each of these paths were traced from a starting point at 1.1~$R_{\sun}$. We highlight one trajectory on the lower left in blue and we use this path as an example throughout the paper in describing our methods.}
	\label{fig:wavetrack}
\end{figure}

This process was repeated for every pixel in the COMP data to build up a map of the wave travel directions. We can then track waves through the map by selecting a starting pixel and then using Euler's method to step forward and backward along the path to solve for the entire wave trajectory. Figure~\ref{fig:wavetrack} shows examples of wave trajectories traced starting from initial positions located at a radius of $r=1.1$~$R_{\sun}$ and varying polar angles, $\theta$. For clarity, the plot illustrates wavetracks starting only from every fifth pixel along that arc. The resulting wave trajectories show good agreement with the apparent shapes of coronal structures seen in Figures~\ref{fig:context} and \ref{fig:contexthalf}, which is evidence that the method is accurately tracking the waves. One exception is at the west limb, near the active regions, where the wave trajectories take on a jagged appearance due to the complex intervening structures along the line of sight. 

We will compare our results to several other publications also based on COMP, so it would be useful to discuss some of the differences between the wave tracking method used here and those used by others. Our method is most similar to that of \citet{Tomczyk:Sci:2007}, except that \citeauthor{Tomczyk:Sci:2007} Fourier filtered the time series using a Gaussian filter centered at 3.5~mHz, whereas we have only used a low pass filter. In a later paper, \citet{Tomczyk:ApJ:2009} used a coherence based method \citep{McIntosh:SolPhys:2008} and the same method was later used by \citet{Sharma:NA:2023}. The coherence between the time series is computed using the cross spectrum. Because the raw cross spectrum of two time series will always give a coherence of unity \citep{Jenkins:book}, it is necessary to average the cross spectra in some way to obtain a meaningful coherence value. For the above studies, this averaging takes place at the step where the Gaussian filter is applied. The reason we have not adopted this approach is that our objective is to understand the relationship between the $\delta v$ and $\delta i$ fluctuations and we posit that this might involve a frequency scaling between the waves. As such, we avoided assuming that the waves have a particular frequency by not filtering the data at 3.5~mHz. A drawback is that noise may be suppressed less here than in those other studies. 

\subsection{Perpendicular correlation lengths}\label{subsec:corlength}

The perpendicular correlation length can be quantified directly from the constructed cross-correlation calculations. This length represents the extent to which the plasma is perturbed in the perpendicular direction due to the wave. It is also related to the size of the source of wave excitation, analogous to the size of the antenna in a laboratory plasma. In the context of turbulence, it may be thought of as the largest scale eddy. 

To quantify this perpendicular correlation length, we found the characteristic transverse width of the island of high correlation by fitting it with a Gaussian. The island of high correlation is peaked in the center, with the central pixel having correlation coefficient $c_{ii}=1$, and declines toward the boundaries. This island can be modeled as a two-dimensional Gaussian distribution \citep[e.g.,][]{Sharma:NA:2023} and the width characterized by the standard deviations of the parallel and perpendicular axes, $\sigma_{\parallel}$ and $\sigma_{\perp}$. To determine the value of $\sigma_{\perp}$, we summed the $c_{ij}$ along the parallel direction and fit the resulting distribution of $c_{ij}$ versus perpendicular distance with a one-dimensional Gaussian. Since the integral of a two-dimensional Gaussian along one axis preserves the $\sigma$ for the other axis, this summation reduces the dimensionality of the problem without affecting $\sigma_{\perp}$. \citet{Sharma:NA:2023} noted that solar wind studies of perpendicular correlation lengths report the exponential $1/e$ length scales given by an autocorrelation function; and so in order to better compare to these values, we likewise scale our Gaussian correlation lengths by a factor of $\sqrt{2}$ defining $L_{\perp}=\sqrt{2}\sigma_{\perp}$. 

The perpendicular correlation lengths for $\delta v$ take a broad range of values and a histogram resembles a lognormal distribution, as has been found by \citet{Sharma:NA:2023}. We found the median $L_{\perp} = 7.0$~Mm and the most probable value (i.e., the mode) is $L_{\perp} = 5.0$~Mm. This is somewhat smaller than found by \citet{Sharma:NA:2023} who found that the most probable values were in the range of $L_{\perp} = 7.6-9.3$~Mm among several observations. One possible reason for the discrepancy is that \citeauthor{Sharma:NA:2023} focuses specifically on fluctuations at frequencies of 3.5~mHz, whereas our data is less strongly filtered. Given the broad statistical distribution of $L_{\perp}$ values in the measurements, there is still substantial overlap in these measurements and the discrepancy is small. 

\begin{figure}[h]
	\centering \includegraphics[width=0.8\textwidth]{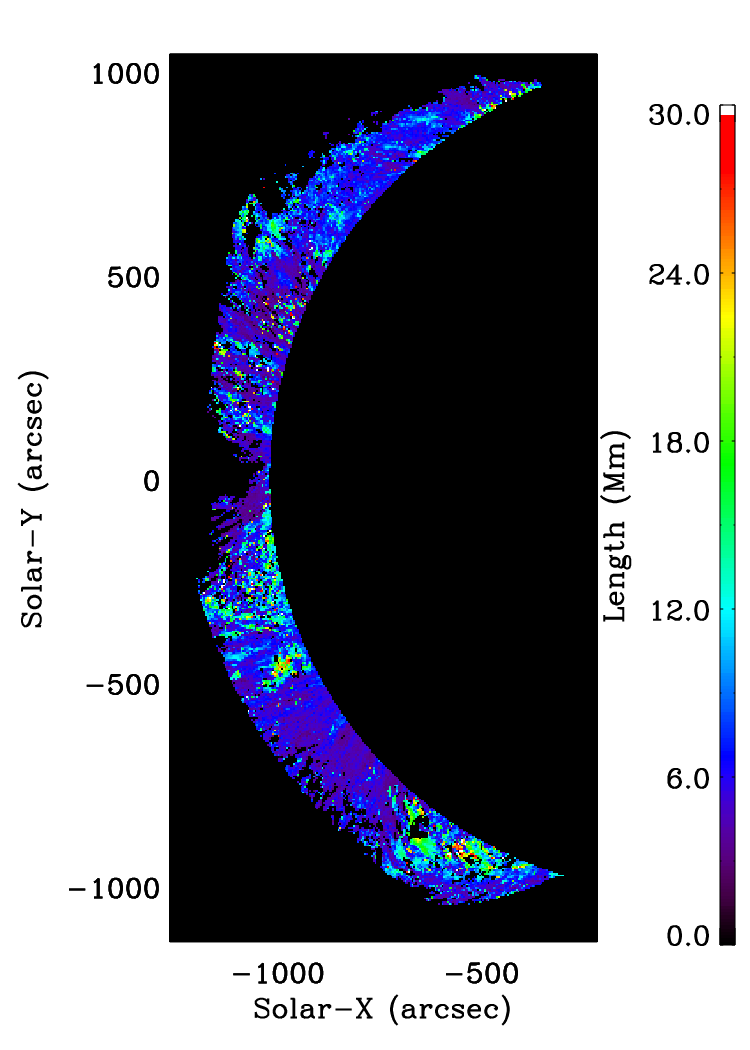}
	\caption{Map of $L_{\perp}$ for the $\delta v$ fluctuations.}
	\label{fig:vlper} 
\end{figure}

The $L_{\perp}$ map derived from $\delta v$ (Figure~\ref{fig:vlper}) shows relatively low $L_{\perp}$ areas ($L_{\perp} \lesssim 9$~Mm) that appear to follow the coronal loops in the images. There are clumps of high $L_{\perp}$, particularly at low heights and also the active regions on the west limb (not shown in the figure). One possible interpretation is that the low $L_{\perp}$ represents Alfv\'enic waves traveling along individual magnetic flux tubes, while the high $L_{\perp}$ represents structures undergoing large scale oscillations so that $L_{\perp}$ is due to the entire structure being generally excited rather than to characteristics of a traveling wave mode. We found no discernable trend of $L_{\perp}$ varying with height. 

\begin{figure}[h]
	\centering \includegraphics[width=0.8\textwidth]{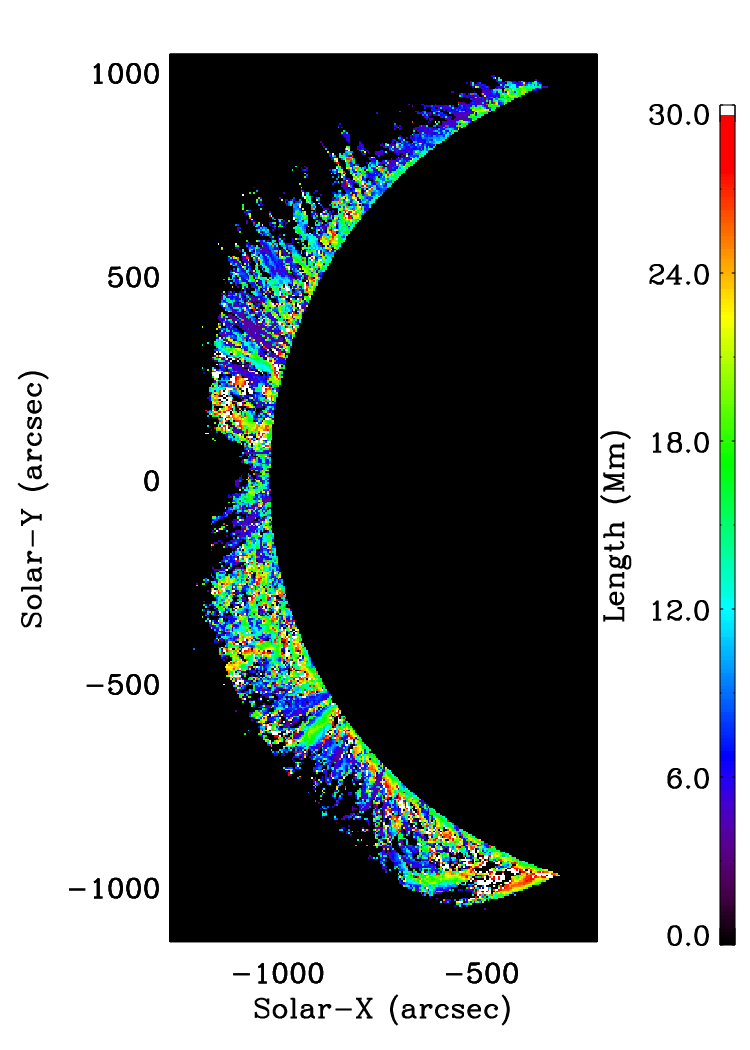}
	\caption{Map of $L_{\perp}$ for the $\delta i$ fluctuations.}
	\label{fig:ilper} 
\end{figure}

With respect to intensity fluctuations (Figure~\ref{fig:ilper}), the range of $L_{\perp}$ derived from $\delta i$ was even broader than that for $\delta v$, having a median value of $L_{\perp} = 10.8$~Mm and a mode of $L_{\perp} = 3.7$~Mm, though still comparable to the range found for the $\delta v$. The spatial distribution for the $\delta i$ $L_{\perp}$ shows broader structures than that for $\delta v$. These may represent regions affected by large scale intensity changes as those structures undergo heating and cooling throughout. Those processes may not lead to similar signatures in the Doppler velocities, because any associated flows are likely to be parallel to the magnetic field and so perpendicular to the line of sight at the limb, resulting in minimal Doppler shifts. 

It is interesting that $L_{\perp}$ for both $\delta v$ and $\delta i$ fluctuations are an order of magnitude larger than apparent widths of coronal loops. Analysis of loops in high-spatial-resolution images has shown that the typical widths of coronal loops are $\approx 0.5$--1.5~Mm \citep{Peter:AA:2013, Aschwanden:ApJ:2017}. One might have expected that phase mixing at the boundary of coronal loops would interrupt the correlations leading to wave patterns and loops having similar size distributions. One possible reason this does not occur comes from simulations of Alv\'en waves on multi-stranded loops showing that strands can influence their neighbors, which would result in $L_{\perp}$ larger than the scale of the underlying structures \citep{Guo:ApJ:2019}. Another possibility is that the images of loops may not be an accurate representation of the true coronal density structure \citep[e.g.,][]{Malanushenko:ApJ:2022}. 

Similarly, we also expect to find small values of $L_{\perp}$ if the fluctuations are related turbulence. Turbulence is thought to be important for the generation and evolution of Alfv\'enic waves in the solar corona. If the fluid motions that generate the Alfv\'enic waves are turbulent, we would expect waves to be excited with a broad range of perpendicular length scales. Moreover, if turbulence occurs in the corona, then we would expect energy initially excited at large $L_{\perp}$ to cascade to small $L_{\perp}$. A potential physical explanation for why small values of $L_{\perp}$ are not observed is that small values of $L_{\perp}$ are efficiently dissipated. 

It is possible, though, that we observe few small values of $L_{\perp}$ because the measurements have limitations from spatial resolution, noise, and line-of-sight averaging. To test this, we have studied in more detail the distribution of our $L_{\perp}$ measurements. \citet{Sharma:NA:2023} commented that the distribution of $L_{\perp}$ values in the COMP data appear to approximately follow a lognormal distribution and we find a similar distribution here. In Appendix~\ref{app:distribution} we investigate the statistical distribution in more detail with the aim to determine whether there may be small $L_{\perp}$ values that are not resolved. In particular, we consider the hypothesis that the actual distribution of $L_{\perp}$ in the corona follows a power law, and argue that the addition of noise and line-of-sight averaging can distort the measurements, producing a histogram that can match the observed one. The implication is that known systematic issues are capable of biasing our $L_{\perp}$ measurements to larger values. However, the arguments are speculative (which is why we relegate them to the Appendix).

Future work to understand the perpendicular correlation lengths will require higher spatial resolution measurements and a deeper understanding of systematic uncertainties in the observations. We plan to analyze measurements with higher spatial resolution instruments, such as the Daniel K. Inouye Solar Telescope \citep[DKIST;][]{Rast:SolPhys:2021} to address some of these limitations. 

\clearpage
\subsection{Time-distance plots and $k$-$\omega$ diagrams}\label{subsec:kw}

We can measure properties of the wave as they propagate along the trajectories identified in Figure~\ref{fig:wavetrack}, allowing us to infer the wave speed and direction. To do this, we first combine the time series of fluctuations at each pixel along the wave path to construct time distance plots $\delta i (t,s)$ and $\delta v (t,s)$, where $s$ is the distance along the trajectory. 

\begin{figure}[h]
	\centering \includegraphics[width=0.8\textwidth]{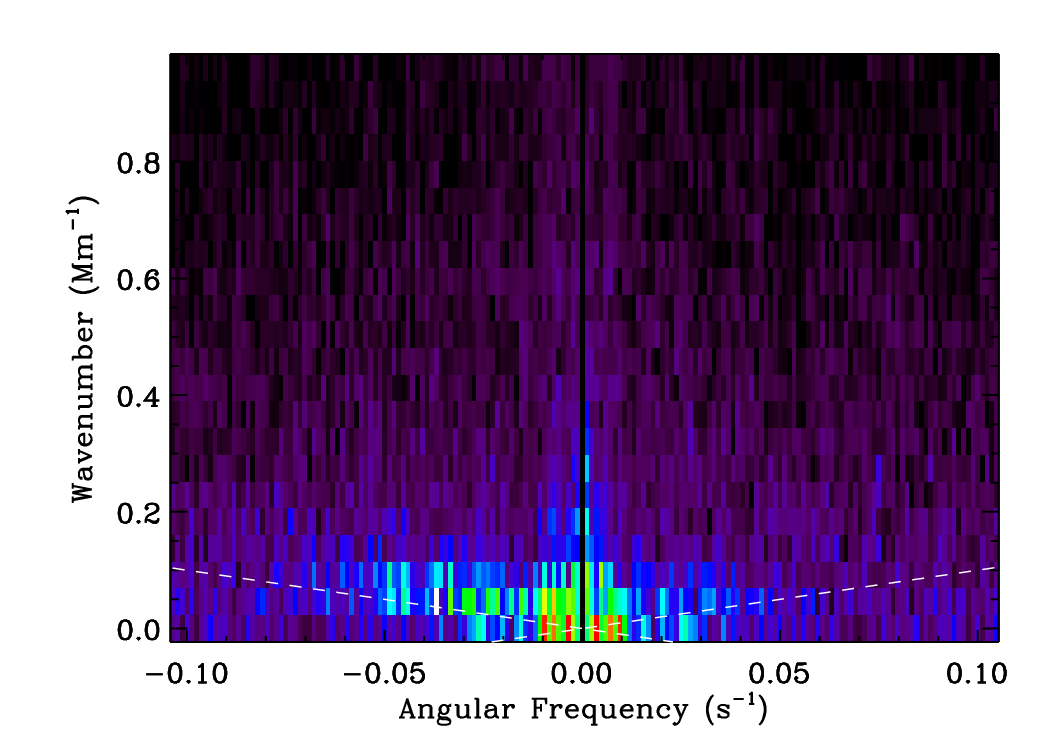}
	\caption{$k$-$\omega$ diagram for the $\delta v$ fluctuations along the trajectory shown in Figure~\ref{fig:wavetrack}. The dashed lines indicate a wave speed of 1~$\mathrm{Mm\,s^{-1}}$, which is a typical Alfv\'en speed in the corona.}
	\label{fig:vkw} 
\end{figure}

Next, we took the Fourier transform of the time-distance plot and constructed a wavenumber-frequency $k$-$\omega$ diagram. Note that it is conventional to use angular frequency $\omega$ for such a plot, even though elsewhere we usually refer to the oscillation frequency $f$. Figure~\ref{fig:vkw} shows the $k$-$\omega$ diagram for $\delta v$ fluctuations along the wave trace indicated in Figure~\ref{fig:wavetrack}. Here we have used the sign convention that negative $\omega$ corresponds to antisunward (upward) traveling waves and positive $\omega$ corresponds to sunward (downward) traveling waves. 

In Figure~\ref{fig:vkw} the wave power appears concentrated along two ridges. Since the phase speed of a wave is given by $\omega/k$, these ridges have a slope that is the inverse of the wave phase speed. The dashed lines in the figure correspond to 1~$\mathrm{Mm\,s^{-1}}$, which is a typical Alfv\'en speed in the corona. The dashed lines were drawn by hand to ``guide the eye'', yet it does align roughly with a ridge in the power spectrum, suggesting that the $\delta v$ fluctuations are due to Alfv\'enic waves, such as kink or torsional Alfv\'en modes. More precise measurements of the phase speed are presented in Section~\ref{subsec:vel}. 

\begin{figure}[h]
	\centering \includegraphics[width=0.8\textwidth]{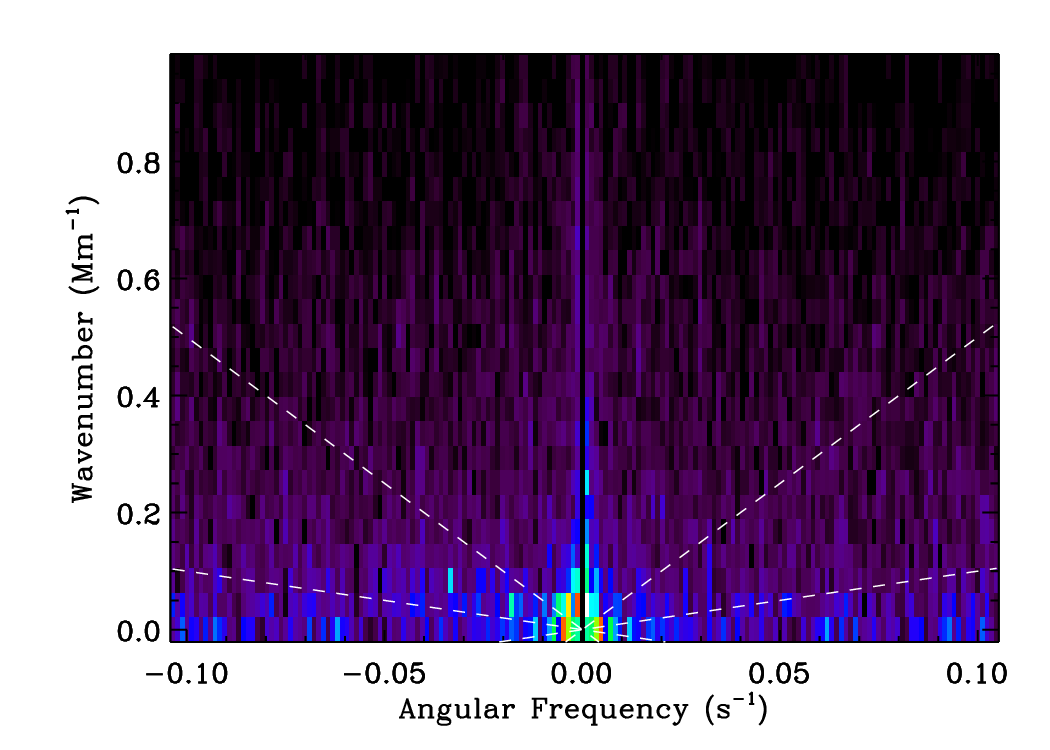}
	\caption{$k$-$\omega$ diagram for the $\delta i$ fluctuations along the trajectory shown in Figure~\ref{fig:wavetrack}. The dashed lines indicate wave speed of 1 and 0.2~$\mathrm{Mm\,s^{-1}}$, corresponding to typical Alfv\'en and sound speeds in the corona.}
	\label{fig:ikw} 
\end{figure}

Figure~\ref{fig:ikw} plots the $k-\omega$ diagram derived for $\delta i$ fluctuations along the same trajectory.  Dashed lines on the plot correspond to wave speeds of 1~$\mathrm{Mm\,s^{-1}}$ and 0.2~$\mathrm{Mm\,s^{-1}}$, representing typical Alfv\'en and sound speeds, respectively. Although there is some $\delta i$ power along and between these lines, we do not find distinct ridges indicative of a particular wave mode. The majority of the power is concentrated at low $\omega$ and low $k$, which likely represents long term changes in the intensity of large scale structures due to heating, cooling, or other evolution. That the power does not align with a clear wave mode suggests that much of the $\delta i$ variation observed in the corona is not due to waves, although waves are likely also present.

\subsection{Upward versus downward waves}\label{subsec:updown}

We can estimate the balance of upward versus downward waves by comparing the total power $P$ at negative versus positive frequencies, respectively, having finite wavenumber $k > 0$. We will quantify this ratio by defining $\eta = (P_{\mathrm{up}}-P_{\mathrm{down}})/(P_{\mathrm{up}}+P_{\mathrm{down}})$. Because noise power will affect the denominator in $\eta$, we first estimate the noise level and subtract it from each pixel. The $k-\omega$ diagram suggests that the wave power is concentrated in the ridges at low $k$ and that the power at high $k$ and high $\omega$ is likely due to noise. So, we estimate the noise in each $k$ and $\omega$ bin as being the average power in upper corners of the diagram containing the highest $10\%$ of wavenumbers and angular frequencies. This value is then subtracted from each bin of the diagram before computing $\eta$. For the example shown in Figure~\ref{fig:vkw}, we find $\eta_{\delta v} = 0.24$, indicating predominantly upward-propagating waves. We compute the imbalance of upward versus downward $\delta i$ power in the same way. Similarly, for the $\delta i$ fluctuations shown in Figure~\ref{fig:ikw}, we find $\eta_{\delta i} = 0.07$, indicating that traveling $\delta i$ fluctuations tend to be moving upward. The magnitude of $\eta_{\delta i}$ is smaller than $\eta_{\delta v}$, but this is likely because a significant amount of the $\delta i$ fluctuation power is not in a traveling wave. 

\begin{figure}[t]
	\centering \includegraphics[width=1.0\textwidth, trim={2cm 0 2cm 0}]{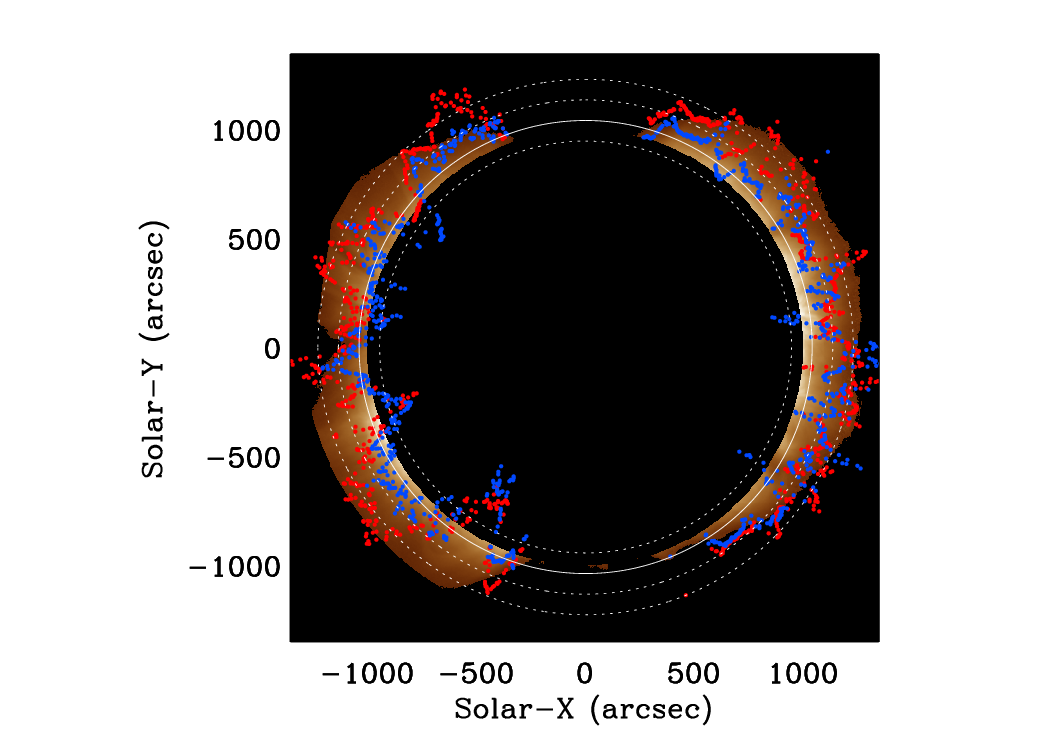}
	\caption{Polar plot illustrating $\eta_{\delta v}$ (red symbols) and $\eta_{\delta i}$ (blue symbols) for wave trajectories at varying polar angles with starting points at 1.1~$R_{\sun}$. The solid circle indicates $\eta = 0$ with larger radii corresponding to $\eta > 0$ and smaller radii $\eta < 0$. Dotted circles are drawn at intervals of $\Delta \eta = 0.1$ for scale. The background image is the average Fe~\textsc{xiii} intensity as in Figure~\ref{fig:context}.} 
	\label{fig:updown} 
\end{figure}

Figure~\ref{fig:updown} shows a polar plot illustrating the wave power imbalance for all wave trajectories starting at $1.1$~$R_{\sun}$ like those in Figure~\ref{fig:wavetrack}. Red symbols indicate the imbalance for velocity fluctuations, $\eta_{\delta v}$, and blue symbols for intensity fluctuations, $\eta_{\delta i}$. The solid circle indicates zero, so symbols at larger radii indicate predominantly upward traveling waves while those at smaller radii indicate predominantly downward traveling waves. For scale, the dotted circles are drawn at intervals of $\Delta \eta = 0.1$. 

We find that $\eta_{\delta v}$ and $\eta_{\delta i}$ are usually positive. Although there appear to be exceptions at several locations, referring to Figure~\ref{fig:wavetrack} shows that these locations all correspond to trajectories that travel along the top of a short coronal loop. In such cases our definition of ``up'' and ``down'' along the wave trajectory does not have a clear interpretation as being towards or away from the Sun. However, it is interesting that such loops do show a clear imbalance. One possibility is that this indicates a stronger wave driving from one footpoint of the loop compared to the other. We also find that $\eta_{\delta v}$ and $\eta_{\delta i}$ are strongly correlated with one another. Quantitatively, their correlation has a Spearman coefficient of $0.8$. We find average magnitudes of $\eta_{\delta v} = 0.14$ and $\eta_{\delta i} = 0.09$, indicating waves are traveling predominantly, though not exclusively, away from the Sun. The smaller value for $\eta_{\delta i}$ is likely due to $\delta i$ having relatively more power in fluctuations that are not traveling compared to $\delta v$. 


\subsection{Wave speeds and Plasma Beta}\label{subsec:vel}

\begin{figure}[t]
	\centering \includegraphics[width=0.8\textwidth]{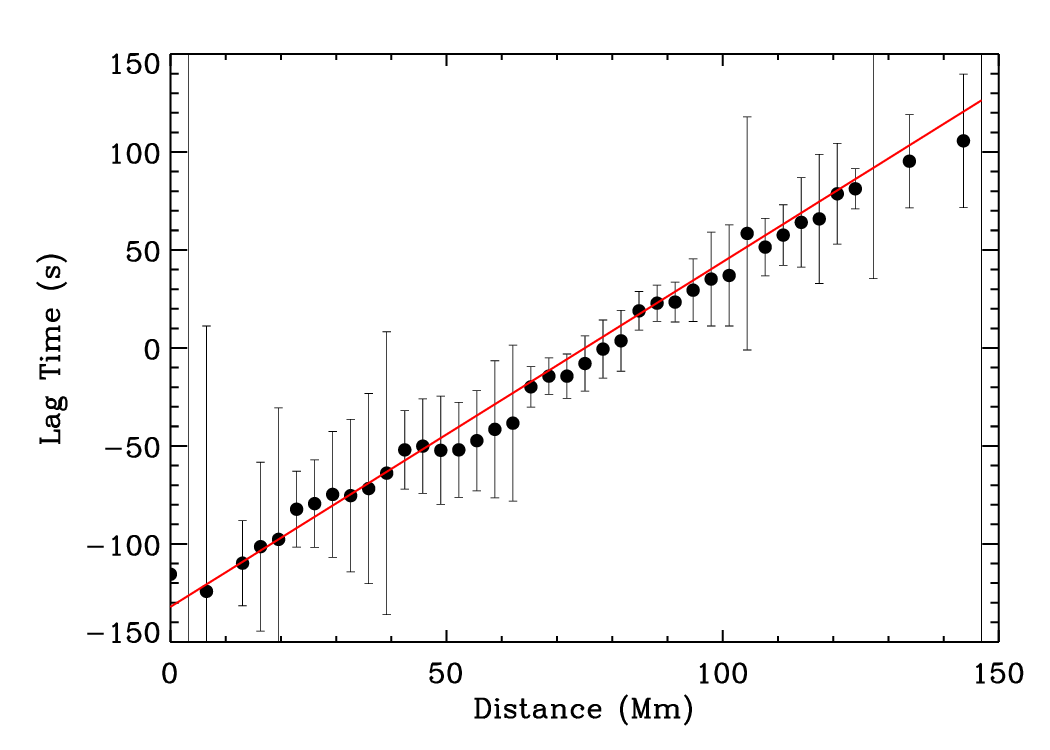}
	\caption{Lag time versus distance for $\delta v$ fluctuations along the trajectory highlighted in Figure~\ref{fig:wavetrack}. The vertical error bars are the uncertainties in the lag time from the parabolic fit to the cross correlation $c_{ij}(\tau)$.} 
	\label{fig:vspeedex} 
\end{figure}

We have determined the phase speeds of the waves along the inferred trajectories by separating the upward versus downward waves and using a cross-correlation method to determine the lag time between different positions along the path, following \citet{Tomczyk:ApJ:2009}. The separation of the upward versus downward waves is necessary because the superposition of counterpropagating waves of equal amplitude will form a standing wave, which the cross-correlation method will see as having an infinite phase speed. Even if the amplitudes are not equal, the phase speed will be biased. Thus, the first step for determining the phase speeds is to construct a time-distance map that selects only one direction of waves. We do this by performing the inverse Fourier transform of the $k-\omega$ diagram using either only the negative (outward) or positive (inward) wave $\omega$. Next we select a reference height $i$ near the center of the wave trajectory. We then find the cross correlation of the fluctuations at each other height $j$ with the reference as a function of lag time $c_{ij}(\tau)$. The cross correlation as a function of the lag time is fit with a parabola to find the lag time $\tau_{ij}$ that maximizes $c_{ij}$. Finally, we fit a linear function to the plot of lag times $\tau_{ij}$ versus distance from the reference pixel $\Delta s_{ij}$ and obtain the speed as the slope of that line. Figure~\ref{fig:vspeedex} shows an example of such a fit for upward $\delta v$ fluctuations along the trajectory highlighted in Figure~\ref{fig:wavetrack}. Note that this method is only sensitive to the components of the velocity in the plane-of-sky. 

We applied the lag time analysis to all of the wave trajectories that pass through 1.1~$R_{\sun}$. For both $\delta i$ and $\delta v$ fluctuations, there was no significant difference in the phase speeds in the upward versus downward direction. This makes sense as the mean flow speeds at these low heights are expected to be small so there is no physical reason for the measured fluctuation speeds to differ. As we have earlier shown that there is more power in the upward fluctuations, the upward signals are relatively less affected by noise and their speeds are typically measured with greater precision. For that reason, the rest of the discussion here will be based only on the upward fluctuation speeds. 

\begin{figure}[h]
	\centering \includegraphics[width=1.0\textwidth, trim={2cm 0 2cm 0}]{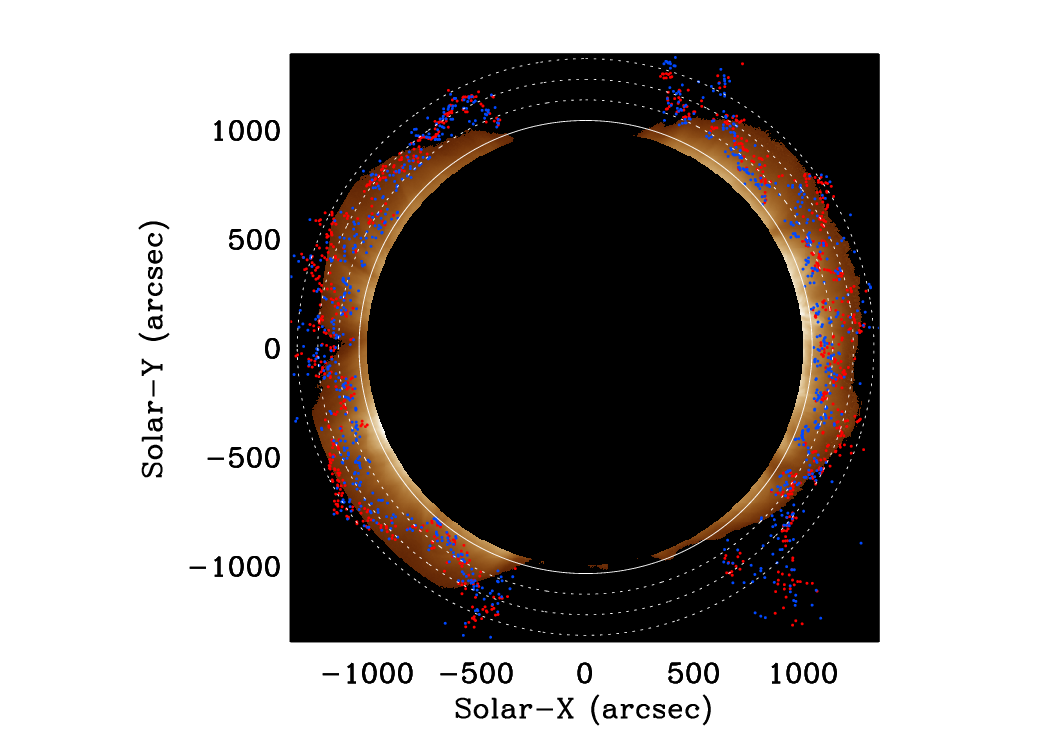}
	\caption{Plane-of-sky propagation speeds for $\delta v$ fluctuations (red symbols) and $\delta i$ fluctuations (blue symbols). The speed is indicated by the radial distance from the solid circle, which corresponds to a speed of zero. Dotted circles are at intervals of $200$~$\mathrm{km\,s^{-1}}$. } 
	\label{fig:speed} 
\end{figure}

Figure~\ref{fig:speed} shows a polar plot of the measured wave speeds. The red symbols correspond to $\delta v$ fluctuations and the blue symbols to $\delta i$ fluctuations. The speed is proportional to the radial distance from the solid circle, which represents zero, and the dotted circles indicate intervals of 200~$\mathrm{km\,s^{-1}}$. Because we are measuring the motions of the fluctuations as they propagate through the image, these speeds correspond only to the plane-of-sky velocity. The actual propagation speeds are necessarily greater than those we have inferred because we are ignorant of the line-of-sight component of the speed. However, as $\delta v$ and $\delta i$ appear to propagate in the same direction, it remains valid to compare their speeds to one another. 

The relationship between the speeds of $\delta v$ versus $\delta i$ fluctuations appears to fall into two classes depending on whether the fluctuations are traveling along short closed loops or long paths that do not close within the bounds of our measurements (refer to Figure~\ref{fig:wavetrack}). The short closed loops are found at middle latitudes. In these locations, the speeds of the $\delta v$ and $\delta i$ fluctuations are nearly the same. This suggests that both types of fluctuations in these locations reflect the same underlying wave mode, such as compressive modes along these loops. 

The long open trajectories, seen most clearly at moderately low latitudes on the east limb, exhibit significant differences in the wave speeds of $\delta v$ versus $\delta i$ fluctuations. The speeds of the $\delta v$ fluctuations are greater by several hundred $\mathrm{km\,s^{-1}}$ than those of the $\delta i$ fluctuations. This suggests that the $\delta v$ fluctuations here correspond to Alfv\'enic waves and the $\delta i$ fluctuations are dominated by slow mode acoustic waves. If we adopt this interpretation, then we may estimate the plasma $\beta$, i.e., the ratio of the fluid to magnetic pressure, from the ratio of the observed sound and Alfv\'en speeds as $\beta \sim c_{\mathrm{s}}^2/V_{\mathrm{A}}^2$. The median value of $\beta$ for these open trajectories is then about $\beta \approx 0.5$. 

\section{Discussion}\label{sec:dis}

The $\delta v$ fluctuations observed in the corona appear to be predominantly in the form of Alfv\'enic waves, as has been previously recognized \citep{Tomczyk:Sci:2007}. Fourier analysis of the time-distance maps of these waves (Figure~\ref{fig:vkw}) suggests that they obey a linear dispersion relation with propagation speeds of $\gtrsim 500$~$\mathrm{km\,s^{-1}}$, which are consistent with Alfv\'enic waves.

The physical significance of the $\delta i$ fluctuations is less clear. They likely arise due to multiple phenomena. In some regions, especially relatively short closed loops, the dominant $\delta i$ fluctuations appear to have similar speeds and frequencies to the $\delta v$ fluctuations. In these cases, $\delta i$ may be due to density variations associated with fast compressional modes. In other regions, corresponding usually to long trajectories, we find that the $\delta i$ fluctuations propagate slower than $\delta v$ fluctuations and appear at different frequencies. In these cases, much of the $\delta i$ fluctuation power may come from slow mode acoustic waves. However, even for these open trajectories, the $k$-$\omega$ diagrams for $\delta i$ fluctuations do not resolve into a clear dispersion relation (e.g., Figure~\ref{fig:ikw}). In all regions of the corona, a significant amount of density fluctuation power is likely due to erratic changes that are not associated with waves. Such sources may include eruptions, flows, and impulsive heating events. This ambiguity adds to the well-known problem of distinguishing slow mode waves from quasi-periodic pulsations \citep[see][and references therein]{Banerjee:SSR:2021}. 

In describing these characteristics, we have noted that there is an apparent distinction between the behavior of fluctuations on short loops and open trajectories. By short loops we mean, essentially, those whose maximum height is around 1.3~$R_{\sun}$ so that the top of the loop falls within our field of view. The long trajectories are those regions where the fluctuations appear to propagate roughly radially from the base of the corona to the outer radius of our available data. It is interesting to note that these open trajectories do not correspond to open magnetic field lines. Rather, they appear to correspond to the bases of long loops whose loop tops are well beyond the field of view of the COMP data. Both of these structures are expected to have plasma properties typical of the quiet Sun, and so the apparent differences in the behavior of fluctuations along these structures is likely related to the loop length. 

Much previous work has distinguished between the behavior of waves on open field lines versus closed field regions, but our observations suggest that there is an intermediate behavior on long loops. The reasons for this are not yet clear and should be investigated in future work. However, we can speculate on several explanations. One possibility is that the behavior is related to the ability to form standing modes on the loops. In this case, it may be that the short loops are capable of supporting low-order resonances, but for the longer loops the dominant wave frequencies are non-resonant. Another possibility is that the long loops are open from the waves' perspective because of dissipation at large heights. That is, on the shorter loops waves from opposite footpoints still retain significant wave power at the locations where they collide with one another. However, on the long loops, waves from opposite footpoints lose most of their energy before they encounter waves propagating up from the opposite footpoint. From the perspective of the waves, such a long loop would be similar to an open field line. 

Even on the long trajectories, we found that there is a substantial fraction of Alfv\'enic wave energy that is traveling towards the Sun, as can be seen from the positive frequency half of the $k$-$\omega$ diagrams. Our estimated average $\eta_{\delta v} = 0.14$ corresponds to up to 40\% of the wave energy traveling towards the Sun. However, this is likely an overestimate because it is sensitive to noise level in estimating the normalization for $\eta$. If it is the case that the injected wave energy from the footpoints is dissipated before it can reach the other half of the loop, then the downward wave energy must be generated within the loop as a reflection or instability of the incident upward propagating waves. Many of the possible processes that would cause this involve interactions between the Alfv\'enic waves and density fluctuations, either because the density fluctuations enhance the wave reflection or due to an interaction through PDI. 

One objective of this investigation was to determine the nature of the relationship between the $\delta v$ and $\delta i$ fluctuations. We were particularly interested in seeing whether PDI is a common process in the corona, but the data here are ambiguous. This is largely because the $\delta i$ fluctuations appear to be caused by several different physical processes, which impedes our ability to make a direct comparison with $\delta v$ fluctuations such as was done in \citet{Hahn:ApJ:2022}. Tentatively, it appears that PDI is unimportant on the shorter closed loops, whose dynamics we hypothesize to be more dominated by Alfv\'enic wave collisions between the primary Alfv\'enic waves traveling up from the footpoints. \rev{This is consistent with the relevant timescales as well. The Alfv\'en waves have periods of about 5~minutes and for the measured Alfv\'en speed the travel time from one footpoint to the other of a short $\sim 500$~Mm loop is about 20~minutes, or a few wave periods. The PDI growth rate is expected to be a small fraction of the pump wave frequency \citep[see e.g.,][]{Hahn:ApJ:2022}, leading to growth timescales of hours. Thus, Alfv\'en wave collisions are probably a more important effect than PDI on short loops. In contrast,} there do appear to be frequency scaling relations on the longer wave trajectories that are suggestive of PDI, which would very likely also apply to open field regions. Future measurements should aim to verify this by focusing on open structures. 

Additionally, our analysis of the cross-power spectrum showed that there are structures where the $\delta i$ and $\delta v$ fluctuations at a given frequency show a definite phase difference. This suggests that they arise from the same underlying wave mode. Although there is much scatter in the data, the dichotomy between long and short loops appears also to be true in these measurements, where there is a tendency for the longer loops to exhibit phase differences of $\phi \sim \pi/2$ versus short loops having phase differences near zero.

\section{Summary}\label{sec:sum}

We have investigated the general properties of $\delta v$ and $\delta i$ fluctuations in the quiet Sun corona and their interactions with one another. The $\delta v$ fluctuations have a distinct dispersion relation in a $k-\omega$ diagram indicating that the majority of those $\delta v$ fluctuations are Alfv\'enic waves. We find typical amplitudes of $0.5$--$1$~$\mathrm{km\,s^{-1}}$ growing with height, at least for the observed heights of $1.1$--$1.3$~$R_{\sun}$. The $\delta v$ fluctuations have median perpendicular correlation lengths of $\approx 7$~Mm. They are predominantly in the outward direction with typically $\eta_{\delta v}\approx 0.14$. These values are in agreement with previous results for $\delta v$ fluctuations observed by COMP \citep[e.g.,][]{Tomczyk:Sci:2007, Tomczyk:ApJ:2009, Morton:ApJ:2016, Sharma:NA:2023}. 

The $\delta i$ fluctuations are a proxy for density fluctuations, but they appear to arise from several physical processes including possibly compressional fast waves, slow mode acoustic waves, and aperiodic variations. The $\delta i$ fluctuations have typical amplitudes of a few percent, which also grow with height over the observed field of view. They have median perpendicular correlation lengths of about $11$~Mm, though with a broad distribution so that this length is not significantly different from that of the $\delta v$ fluctuations. The $\delta i$ fluctuation are also predominantly in the outward direction with $\eta_{\delta i} \approx 0.1$. However $\eta_{\delta i}$ is correlated with $\eta_{\delta v}$. The apparently smaller magnitude of $\eta_{\delta i}$ compared to $\eta_{\delta v}$ is likely because much of $\delta i$ fluctuation power is not in the form of propagating waves. 

The relations between $\delta v$ and $\delta i$ is complex and differs depending on the underlying coronal structures. On short closed loops, we find that power spectra of both types of fluctuations are very similar and that they have similar propagation speeds. Examining the phase relationship at 3.5~mHz on these loops suggests that they have a phase difference near 0. These properties may indicate that both fluctuations are due to the same underlying wave mode. 

On the long quiet Sun structures oriented radially within our data, the $\delta v$ and $\delta i$ fluctuations have different power spectra, which in some cases can be related by a frequency-scaling factor. They also have significantly different speeds, with the $\delta v$ fluctuations traveling hundreds of $\mathrm{km\,s^{-1}}$ faster than the $\delta i$ fluctuations. These longer structures show evidence for wave reflection or PDI, which we plan to investigate in more detail in future work. We also find evidence for another relationship on the long field lines, where we observed $\delta v$ and $\delta i$ fluctuations at the same frequency with consistent phase differences of roughly $\pi/2$, suggesting they are due to the same wave mode. 

In order to better understand the complex variety of relationships between $\delta v$ and $\delta i$ fluctuations in the corona, future measurements should aim to overcome several limitations of the COMP data. First, the COMP data have rather coarse spatial resolution. A similar analysis should be performed on data with a spatial resolution sufficient to resolve the transverse structure of coronal loops. Second, the COMP data observe an Fe~\textsc{xiii} line, which is formed at relatively high temperatures and is therefore insensitive to coronal holes. Hence, we probably do not observe quiescent open magnetic field lines. Measurements of lines formed at cooler temperatures are needed to study open field structures. We next plan to conduct an analysis of DKIST data to address some of these limitations. 

\begin{acknowledgments}
	
M. H. and D. W. S acknowledge support from NSF AST Grant AST-2005887 and NSF SHINE Grant 2229100. X.F. acknowledge support from NASA under Award No. 80NSSC23K0101, and NSF under Award No. 2229101. We thank Mahboubeh Asgari-Targhi for interesting conversations about this work. 
	
\end{acknowledgments}

\software{IDL, solarsoft \citep{Freeland:SolPhys:1998} }

\facilities{COMP}

\clearpage

\appendix 

\section{Systematic factors influencing the measurement of the perpendicular correlation length}\label{app:distribution}

In Section~\ref{subsec:corlength} we described measurements of $L_{\perp}$ for $\delta v$ and $\delta i$ fluctuations. Here, we focus on the $\delta v$ measurements. To derive $L_{\perp}$, we computed the cross correlation between each pixel and a region of neighboring pixels within a box of $\pm 12$ pixels. We took the Gaussian width of these islands as a measure of the perpendicular correlation length. We obtained a sample of $L_{\perp}$ at $N=45,215$ pixels with converged fits within the off-limb corona. A histogram of this set of $L_{\perp}$ measurements resembles a lognormal distribution characterized by a median value of $L_{\perp}$ = $7.0$~Mm and a most probable value of 5.0~Mm (see Figure~\ref{fig:compare}). \citet{Sharma:NA:2023} performed similar measurements of $L_{\perp}$ for $\delta v$ fluctuations, finding similar typical values for $L_{\perp}$ and remarking that their histograms also resembled a lognormal distribution. 

Here, our objective is to explore whether observational biases may prevent the measurement of small $L_{\perp}$ values. In particular, we identify systematic reasons that an underlying power-law distribution may be distorted by the measurements into a form that is roughly lognormal.

\subsection{Analysis}\label{subsec:anal}

\begin{figure}[t]
	\centering \includegraphics[width=0.8\textwidth]{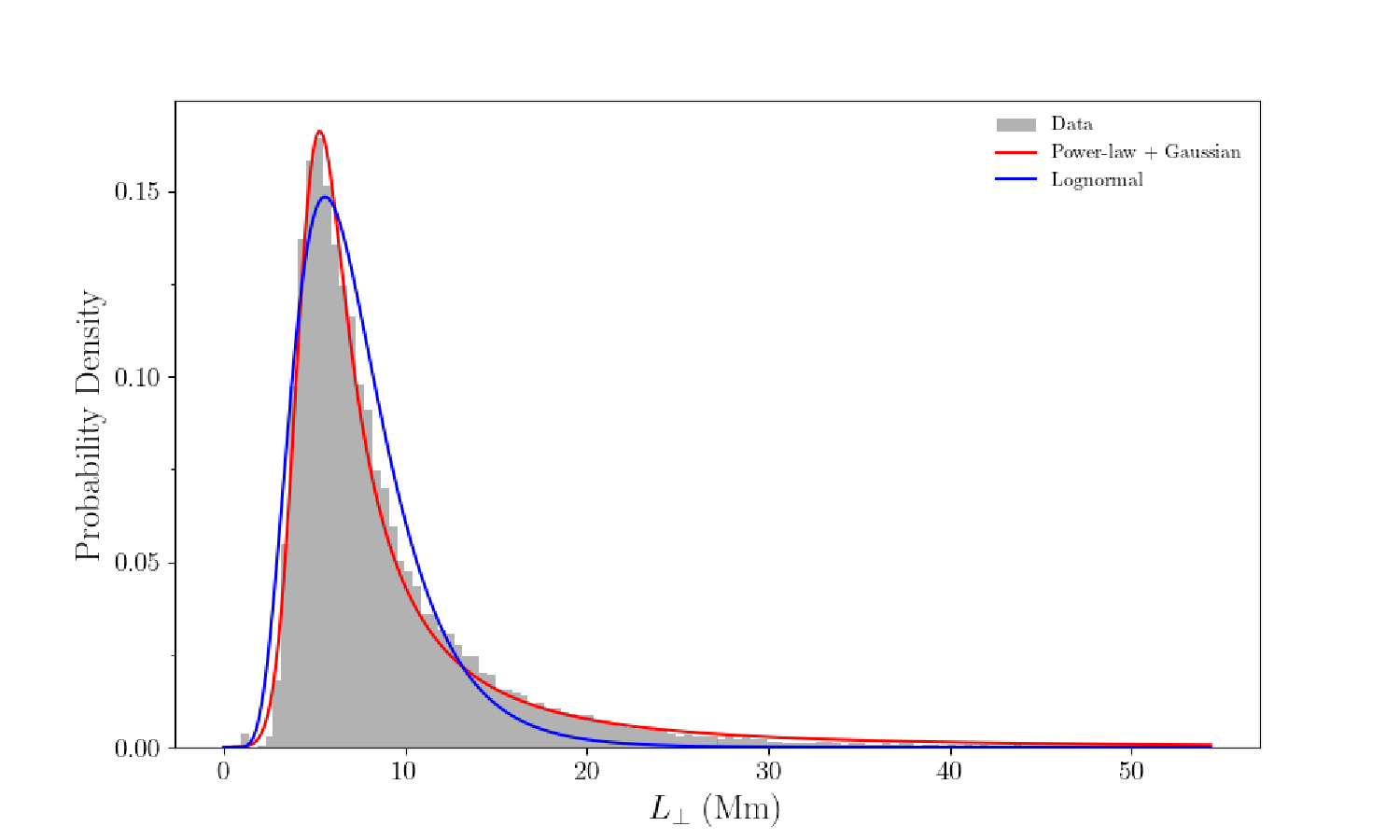}
	\caption{Histogram of the measured $L_{\perp}$ shown as a histogram in grey. The blue curve shows the best-fit lognormal distribution. The red curve illustrates the best-fit distribution obtained by convolving a power law with a Gaussian. The convolved distribution is a better match to the measurements. 
    }
	\label{fig:compare} 
\end{figure}

Figure~\ref{fig:compare} shows a histogram of the $L_{\perp}$ measurements. In constructing the histogram, we omit values larger than 54~Mm, because this $L_{\perp}$ corresponds to a $\sigma_{\perp}$ that exceeds the $\pm 12$ pixel width of field of view used in the correlation analysis described in Section~\ref{subsec:corlength}. We expect fits of such outlier $L_{\perp}$ values to represent poor fits or pixels where the high correlation island was not clear. We then normalize the measured $L_{\perp}$ histogram by dividing by the total number of samples ($N=45,215$) and the width of the binning used for the histogram, in this case $w=0.45$~Mm.
The normalization ensures that the integral of the histogram is equal to one, so that we can compare the distribution to the standard normalized form of probability distribution functions (pdfs). We have verified that the bin size used in constructing the histogram does not qualitatively affect the results and we also make use of metrics that are insensitive to the binning. 

As a first step, we quantify the parameters of the lognormal distribution that best fits the observations. The pdf for the lognormal distribution is given by: 
\begin{equation}
    p(x) = \frac{1}{\sigma x \sqrt{2 \pi}}\exp{\left[-\frac{(\ln(x)-\mu)^2}{2\sigma^2}\right]}. 
    \label{eq:lognormal}
\end{equation}
The parameters $\mu$ and $\sigma$ represent the mean and standard deviation of $\ln(x)$, which follows a normal (Gaussian) distribution. The normal distribution occurs frequently in statistics, because it is a natural consequence of the central limit theorem, namely that the sum of independent random variables tends to follow the normal distribution \citep{Meyer:book}. The lognormal distribution has a similar central limit theorem, except with multiplication rather than addition, because the logarithm of the product is sum of the logarithms of each term. This situation occurs, for example, in fragmentation processes where the size of each fragment of an object is a fraction of its previous size \citep{Crow:book}. 

The blue curve in Figure~\ref{fig:compare} shows the lognormal distribution that best fits the measurements. The best fit parameters were $\mu = 1.91 \pm 0.01$ and $\sigma = 0.439 \pm 0.008$. It would be useful to also describe the perpendicular wavenumber $k_{\perp} \approx 2\pi/L_{\perp}$. The distribution of the inverse of a variable that is lognormally distributed is also lognormal, with the same $\sigma$, but the $\mu$ becomes negative. Accounting for the factor of $2\pi$ in the definition of $k_{\perp}$, the $\mu$ parameter for lognormal $k_{\perp}$ distribution would be $\mu = -0.072$.

\subsubsection{Convolution Model}\label{subsubsec:convolution}

Our hypothesis is that there may be many values with small $L_{\perp}$ that we are not able to measure due to systematic factors. One distribution that is consistent with this hypothesis is a power law. The convolution of a Gaussian and a power-law distribution somewhat resembles a lognormal distribution and can arise if normally distributed random errors are added to power-law distributed data. So, we next investigate whether such a convolution provides a better fit to the measured distribution. The normalized Gaussian distribution with mean zero is given by:
\begin{equation}
    \frac{1}{\sqrt{2\pi}s} \exp{\left[-\frac{x^2}{2s^2}\right]}, 
    \label{eq:gaussian}
\end{equation}
where $s$ represents the standard deviation of the distribution. 
A normalized power-law distribution can be defined described by
\begin{equation}
    P_{\alpha}(x)=\frac{\alpha - 1}{x_0} \left(\frac{x}{x_0}\right)^{-\alpha}, 
    \label{eq:powerlaw}
\end{equation}
where $\alpha > 1$ is the power-law index and $x_0$ is a low-value cutoff, which ensures that the distribution can be normalized. The convolution of these distributions is the integral
\begin{equation}
    P_{\alpha}\star G = \int_{x_0}^\infty P_{\alpha}(x^{\prime})G(x-x^{\prime})\diff{x^{\prime}}. 
    \label{eq:convolve}
\end{equation}
The integration was performed numerically. We will refer to the resulting distribution as the Convolved distribution henceforth. We note that \citet{Parodi:PMB:2006} have obtained closed form solutions in terms of special functions for a different formulation of the power-law distribution that is valid for power-law indices between $0$ and $1$. 

We performed a least-squares fit to the histogram of the $L_{\perp}$ measurements to find the parameters of the best fit convolved distribution parameters. The resulting fit is illustrated by the red curve in Figure~\ref{fig:compare}. The best fit parameters were $\alpha = 2.45 \pm 0.03$, $x_{0} = 4.22 \pm 0.02$~Mm, and $s = 0.97 \pm 0.03$~Mm.  

The convolved distribution is a better fit to the data than the lognormal distribution. This is clear from Figure~\ref{fig:compare}, as well as more quantitative metrics. The least-squares fit minimizes the sum of the squared differences between the observed and model distribution. This value is smaller for the convolved distribution ($1.3\times10^{-3}$) than the lognormal distribution ($6.8\times10^{-3}$). 

We have also used an alternative metric to compare the fits, the Akaike information criterion (AIC), which confirms that the convolved distribution is the better fit. The AIC is computed by first computing the likelihood function $\mathcal{L}$, given by: 
\begin{equation}
    \mathcal{L} = \sum_{i}^{N} \ln{f(x_i)}, 
    \label{eq:likelihood}
\end{equation}
where $f$ is the pdf representing a given model and $x_i$ is the value of the $i$-th measured datapoint, of which there are $N$ in total. The AIC is defined by 
\begin{equation}
    \mathrm{AIC} = 2k - 2 \ln{\mathcal{L}}
    \label{eq:aic}
\end{equation}
where $k$ is the number of parameters in the model: 2 ($\mu$ and $\sigma$) for the lognormal distribution and 3 ($\alpha$, $x_0$, and $s$) for the convolved distribution. A lower value of the AIC indicates a superior fit. In this case, the AIC values are comparable, but favor the Convolved distribution with AIC=$2.5\times10^{5}$ for the Convolved distribution versus $2.6\times10^{5}$ for the lognormal distribution. The AIC has two benefits: First, it uses the data directly and is thereby independent of the binning used in constructing the histogram. Second, it accounts for the different number of parameters in the models. That is, one might expect that a 3-parameter model will give a better fit than a 2-parameter model, simply because there is more freedom to adjust the model. The AIC test accounts for this advantage. 

As a third metric, we have also computed the Kolmogorov-Smirnov (KS) test statistic for both models. The KS test is designed to test whether two sets of data come from the same probability distribution function. It uses the cumulative distribution function, $\mathrm{cdf}(x)$, which is the integral of the pdf up to $x$. The KS statistic is the absolute value of the maximum difference between the cdf of the two distributions being tested. Once again, the KS test shows that the convolved distribution ($\mathrm{KS} = 0.051$) is a better match to the data than the lognormal distribution ($\mathrm{KS} = 0.097$). 

\subsubsection{A simple aggregation model}\label{subsubsec:aggregation}

An alternative model that provides a reasonable match to the observed distribution is that the actual distribution of $L_{\perp}$ follows a power-law distribution, but the measured distribution is distorted by sampling and averaging over that distribution. We will refer to this as the aggregation model. For this model, suppose that the data come from a power-law distribution described by Equation~(\ref{eq:powerlaw}). 
The measured value, though, represents an average of $n$ samples drawn from the power-law distribution with $n$ assumed to follow a Poisson distribution. Physically, this model could represent the averaging of multiple structures along the line of sight in the measurements of the optically-thin corona. 

In order to determine whether this model is consistent with the data and the best-fit parameters, we simulated the process described above. That is, we selected random numbers from a power-law distribution with parameters $\alpha$ and $x_0$ and averaged over $n$ of those numbers, where $n$ was a random number drawn from a Poisson distribution with a mean value of $n_0$. This process was repeated 1,000 times and then the distribution of the resulting averages was examined. In order to identify the best fit parameters ($\alpha$, $x_0$, and $n_0$), we used a differential evolution algorithm to minimize the KS statistic comparing the model distribution to the measurements of $L_{\perp}$. Minimizing the KS metric has several advantages over performing a least-squares fit. One is that the KS metric is independent of the binning for the histogram. More importantly, the KS metric is robust to statistical variations in the model distribution. A least-squares fit requires many more samples to converge, because the samples are generated stochastically and so $\chi^2$ varies with each iteration, even if the fit-parameters remain the same. Because of these random variations, the fit parameters vary slightly with each run, so we estimate the uncertainties on the fit parameters by taking the average and standard deviation of the best fit values over 20 runs. 

\begin{figure}[t]
	\centering \includegraphics[width=0.8\textwidth]{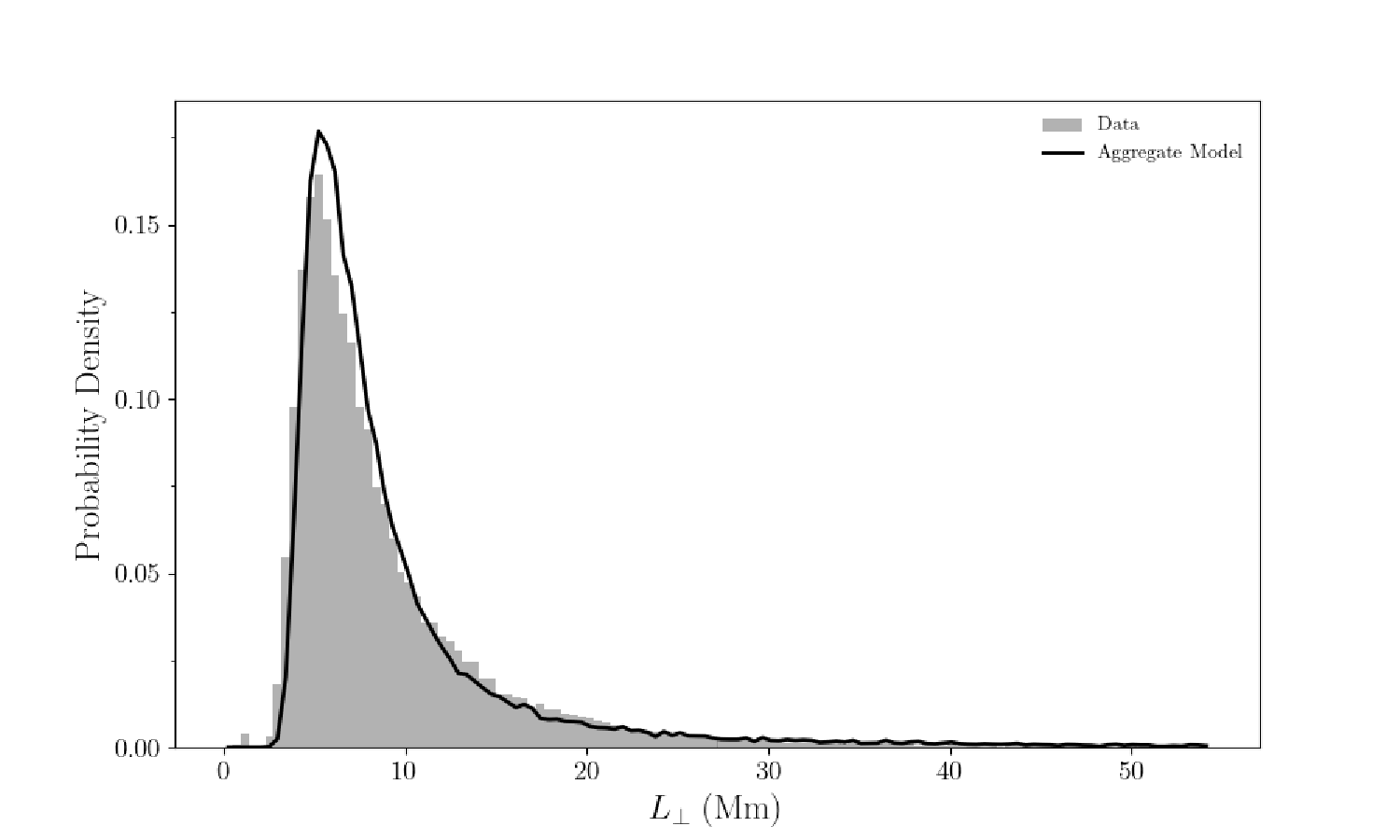}
	\caption{Histogram of the measured $L_{\perp}$ shown as a histogram in grey. The black curve shows the best fit aggregation model, which is obtained by averaging data values drawn from a power-law distribution.
    }
	\label{fig:aggregate} 
\end{figure}

Figure~\ref{fig:aggregate} compares aggregation model to the measurements for the averaged best-fit parameters. These parameters were $\alpha = 2.02 \pm 0.09$, $x_0 = 1.2 \pm 0.4$~Mm, $n = 1.0 \pm 0.3 \times 10^2$. From the figure it is clear that the aggregate model provides a good fit to the data and in fact the average aggregate model KS of $KS=0.046$ was comparable to that of the convolved distribution. It is not possible to compare the other metrics discussed above, because we have not derived an explicit expression for the pdf corresponding to the aggregation model. 

\subsection{Discussion}\label{subsec:disc}

The above models demonstrate that even if the actual distribution of $L_{\perp}$ were statistically distributed as a power law, we could obtain a distribution that approximates a lognormal distribution due to systematic biases in the measurement. One way this can happen is if the measurement involves the addition of normally distributed random noise, which will result in a convolution of the power law with a Gaussian. A second way is if the measurement involves an implicit aggregation or averaging of the power-law distributed values. 

Our measurement process does involve steps that are consistent with these factors. Considering the convolution model, the analysis of the correlation lengths involves computing a cross correlation of time-series data, which has photon counting noise as well as noise from the detector. There are also uncertainties involved in the step where the cross correlation is fit with a Gaussian to infer $\sigma_{\perp}$. From the perspective of the aggregation model, the observed Doppler velocity fluctuations are measured using optically thin emission lines and those Doppler velocities are implicitly averaged along the line of sight. This averaging is known to have a significant effect on the amplitudes of the Doppler fluctuations \citep{McIntosh:ApJ:2012}. Other observational factors may also contribute to the observed distribution. For example, the instrument resolution may limit the ability to measure below a certain length scale. It is suggestive that the convolved fit identifies a cutoff value of $x_0=4.2$~Mm, which is very close to the effective pixel scale 4.5~Mm. There may also be physical limits to the distribution, for example if Alfv\'enic waves with small $L_{\perp}$ are preferentially dissipated. 

The properties of the distribution of $L_{\perp}$ have physical implications. A power-law distribution of $L_{\perp}$ suggests that most of the Alfv\'enic wave energy is concentrated at small $L_{\perp}$ and that large $L_{\perp}$ values occur sporadically. In contrast, a lognormal distribution, suggests that there is a preferred scale for the $L_{\perp}$. For example, there is a peak in the solar p-mode spectrum at spherical harmonics with angular wavenumbers of $l \sim 300$, which corresponds to lengths of $\sim 10$~Mm \citep[e.g.,][]{Hernandez:ApJ:1998}. This would be consistent with the median measured values of $L_{\perp}$ found here. 

In solar physics, the question of distinguishing a power law from a lognormal distribution also occurs in the study of the distribution of flare energies. Many studies of flare energies, or proxies such as the X-ray flux, have inferred that these energies follow a power-law distribution. If $\alpha > 2$ for these energies, then there are sufficiently numerous small events to heat the corona, which is the basis for the nanoflare theory of coronal heating. This power-law form of the nanoflare distribution has been invoked in support of models of nanoflares based on self-organized criticality \citep[][]{Charbonneau:SolPhys:2001} or turbulence \citep[e.g.,][]{Boffetta:PRL:1999}. However, it has also been argued that the flare distribution is not a power law. \citet{Verbeeck:ApJ:2019} shows that some proxies, such as solar flare EUV intensities, are better fit by a lognormal distribution, although other proxies, such as the X-ray flux, do seem consistent with a power law.

Some of the models that we have discussed in terms of the $L_{\perp}$ measurements may have relevance to the problem of understanding the distribution of flares. \citet{DHuys:SolPhys:2016} argued that some of the power-law fits obtained were not performed in a rigorous way. One factor is that many of these fits are performed only for data that exceed a certain threshold. \citet{Buchlin:AA:2005} supported imposing such a threshold in the analysis definition as being consistent with the power-law distribution, because line-of-sight averaging in observational studies causes small events to be not observed. This is essentially the same issue that we have addressed in the aggregation model without imposing an ad hoc threshold. 

Another example in solar physics of the connection between lognormal and power-law distributions concerns the intensity distribution of the quiet Sun and its relationship to nanoflare coronal heating. \citet{Pauluhn:AA:2000} studied the intensity distribution of the quiet Sun chromosphere, transition region, and corona. They found that the distribution of these intensities was well-described by a lognormal distribution. However, nanoflare heating models that assume a power-law energy distribution are able to reproduce these lognormal intensity distributions \citep[][]{Pauluhn:AA:2007, Hahn:ApJ:2022a}. The processes that allow the nanoflare power-law distribution to generate lognormally distributed intensities involve steps that have a similar effect as our convolution and aggregation models. 

\subsection{Summary}\label{subsec:sum}

The statistical distribution of measurements of the perpendicular correlation length of Doppler velocity fluctuations in the corona resembles a lognormal distribution. However, the observations can be better fit by models that consider the underlying distribution to be a power law. These models can represent the addition of noise and the implicit averaging in the observation, along with other systematic factors, such as the influence of spatial resolution. 

These interpretations are tentative. There is insufficient information to determine objectively the true distribution of the data. These $L_{\perp}$ measurements come from different locations throughout the corona. If the characteristics of Alfv\'enic wave excitation are not uniform throughout the corona, structural differences in the corona may cause real variations $L_{\perp}$ and possibly this accounts for the observed distribution. Alternatively, the peak in the $L_{\perp}$ distribution may represent a preferred length scale for injecting wave energy into the corona, such as a peak in the p-mode spectrum. Further insight into the nature of the $L_{\perp}$ measurements will be gained by making measurements with different instruments having different limitations. 

Our conclusion is therefore limited to demonstrating the possibility that existing measurements have not resolved the smallest $L_{\perp}$ values. This possibility has important implications, because it implies that there may be a large population of Alfv\'enic waves in the corona with small perpendicular wavelengths (high $k_{\perp}$). Such waves could arise from the turbulent excitation of the Alfv\'enic waves or Alfv\'enic turbulence in situ driving waves to high $k_{\perp}$. Alternatively, the Alfv\'enic waves might be excited by nanoflares occuring on small unresolved spatial scales. However, our inferred power-law parameters should not be directly interpreted to represent the power spectrum of turbulence as a function of $k_{\perp}$; the assumed $L_{\perp}$ power-law slopes in the opposite direction of that predicted for turbulence, which predicts greater power at larger values of $L_{\perp}$. 
Accurate measurements of the perpendicular correlation lengths or wavenumbers in the corona will require higher resolution measurements and a more comprehensive modeling of systematic factors. 

\bibliography{PDI}
	
\end{document}